\documentclass{emulateapj}

\newcommand{\kms}{km~s$^{-1}$ }
\newcommand{\Lya}{Lyman~$\alpha$~}

\usepackage{lscape} 
\usepackage{color} 

\shorttitle{The Impact of Starbursts on the CGM}
\shortauthors{Heckman et al.}

\begin{document}
\title{COS-Burst: Observations of the Impact of Starburst-Driven Winds on the Properties of the Circum-Galactic Medium}
\author{Timothy Heckman}
\email {theckma1@jhu.edu}
\affil{Center for Astrophysical Sciences, Department of Physics \& Astronomy, Johns Hopkins University, Baltimore, MD 21218, USA}

\author{Sanchayeeta Borthakur}
\affil{Center for Astrophysical Sciences, Department of Physics \& Astronomy, Johns Hopkins University, Baltimore, MD, 21218 USA}

\author{Vivienne Wild}
\affil{ School of Physics and Astronomy, University of St Andrews, St Andrews KY16 9AJ, UK}

\author{David Schiminovich}
\affil{Department of Astronomy, Columbia University, New York, NY 10027, USA}

\author{Rongmon Bordoloi}
\affil{MIT-Kavli Center for Astrophysics and Space Research, 77 Massachusetts Avenue, Cambridge, MA 02139}
\affil{Hubble Fellow}

\begin{abstract}
We report on observations made with the Cosmic Origins Spectrograph (COS) on the Hubble Space Telescope (HST) using background QSOs to probe the circum-galactic medium (CGM) around 17 low-redshift galaxies that are undergoing or have recently undergone a strong starburst (the COS-Burst program). The sightlines extend out to roughly the virial radius of the galaxy halo. We construct control samples of normal star-forming low-redshift galaxies from the COS/HST archive that match the starbursts in terms of galaxy stellar mass and impact parameter.

We find clear evidence that the CGM around the starbursts differs systematically compared to the control galaxies. The Ly$\alpha$, Si III, C IV, and possibly O VI absorption-lines are stronger as a function of impact parameter, and the ratios of the equivalent widths of C IV/Ly$\alpha$ and Si III/Ly$\alpha$ are both larger than in normal star-forming galaxies. We also find that the widths and the velocity offsets (relative to $v_{sys}$) of the Ly$\alpha$ absorption-lines are significantly larger in the CGM of the starbursts, implying velocities of the absorbing material that are roughly twice the halo virial velocity.

We show that these properties can be understood as a consequence of the interaction between a starburst-driven wind and the pre-existing CGM. These results underscore the importance of winds driven from intensely star-forming galaxies in helping drive the evolution of galaxies and the intergalactic medium. They also offer a new probe of the properties of starburst-driven winds and of the CGM itself.  

\end{abstract}

\section{Introduction}

The evolution of galaxies is largely driven by how and when they accrete gas and by how the feedback from newly formed stars and black holes regulates this accretion (see Somerville \& Dav$\acute{e}$ 2015 and references therein).  In turn, the evolution of the inter-galactic medium will be affected by these same feedback processes which can photo-ionize, shock-heat, and chemically-enrich it (e.g. M$\acute{e}$nard et al. 2010). These flows into and out of galaxies will occur within the circum-galactic medium (CGM), a region extending out to roughly the galaxy virial radius. Over the past several years HST/COS observations have greatly improved our understanding of the properties of the CGM in low-z galaxies (e.g. Tumlinson et al. 2013; Stocke et al. 2013; Werk et al. 2014; Liang \& Chen, 2014; Bordoloi et al. 2014a;  Johnson, Chen \& Mulchaey 2015, Borthakur et al. 2015,2016). We now know that the CGM of both star-forming and quiescent galaxies contains a significant reservoir of gas clouds or filaments, most likely at T$\sim 10^4$K and photo-ionized by the diffuse meta-galactic UV background. The CGM in normal star-forming galaxies also contains highly ionized gas traced by O VI that is only rarely present in the CGM of quiescent galaxies (Tumlinson et al. 2011).   

One of the major ways in which feedback occurs is via the outflows of gas driven from strongly star-forming galaxies by the energy and/or momentum injected by massive stars (see Heckman \& Thompson 2017 for a recent review). These galactic winds are ubiquitous in star-forming galaxies at intermediate and high redshift (e.g. Weiner et al. 2009; Steidel et al. 2010; Erb et al. 2012; Kornei et al. 2012; Martin et al. 2012; Bordoloi et al. 2014b; Rubin et al. 2014), while in the present-day universe winds are only observed from starburst galaxies - objects with high star-formation rates per unit area and/or per unit mass (e.g. Heckman et al. 2015; Chisholm et al. 2015; Heckman \& Borthakur 2016; Ho et al. 2016). In principle, galactic winds can have dramatic effects. They may account for the large relative mass of metals in the IGM, for the evolving mass-metallicity relation for galaxies, for the expulsion of baryons from low-mass dark matter halos, and for the transport of low-angular momentum material from forming galaxies (Somerville \& Dav$\acute{e}$ 2015 and references therein).

Unfortunately, these effects have yet to be robustly quantified through direct observation. The principal problems are: 1) The winds are complex multiphase flows whose physical properties and effects on their surroundings can be fully probed through the detailed multi-waveband observations that are possible only in local galaxies. 2) Observations of such outflows from local star-forming galaxies have generally been limited to regions either inside the main body of the galaxy or in the inner-most parts of the CGM (ten-kpc-scale or less). Given that the directly measured outflow speeds in these regions are typically comparable to the galaxy escape velocity, it is not clear whether or how these outflows affect the bulk of the CGM.
  
To date, there has been only a small amount of data testing whether galactic winds from starburst galaxies at low-redshift could have a major impact on the CGM. In Borthakur et al. (2013; hereafter B13) we used data taken in a pilot program of observations of background QSOs with COS to show that the CGM around a small sample of five low-z starburst and post-starburst (SB/PSB) galaxies often have strong C IV absorption-lines arising in a CGM that extends out to impact parameters of $\sim$ 200 kpc.  Lines this strong are not seen in the outer CGM of low-redshift normal star-forming or quiescent galaxies (e.g. Liang \& Chen 2014). The implied masses and densities of this highly ionized material are similar to what has been inferred for the much cooler photo-ionized clouds/filaments seen in the halos of more typical low-z galaxies. Thus, we argued that these results can be understood as the consequence of a starburst-driven wind that has propagated far out into the CGM, interacting with the pre-existing cooler clouds/filaments.
 
In the present paper, we report on the COS-Burst program: new observations that represent a major improvement on the B13 results. We have significantly expanded the size of the sample of low-z starburst/post-starbursts (from five galaxies to seventeen), allowing us to investigate the wind-CGM interaction in a much more statistically robust way. Second, we can now characterize the properties of the CGM in SB/PSB galaxies using multiple ions (not just C IV). In particular, in six cases we can measure the OVI absorption-lines, extending our probe of the CGM to hotter gas than before. Our larger COS-Burst sample also allows us to explore the radial dependence of the CGM properties on the properties of the SB/PSB galaxies.

\begin{figure}
\includegraphics[trim = 0mm 0mm 0mm 0mm, clip,scale=0.45,angle=-0]{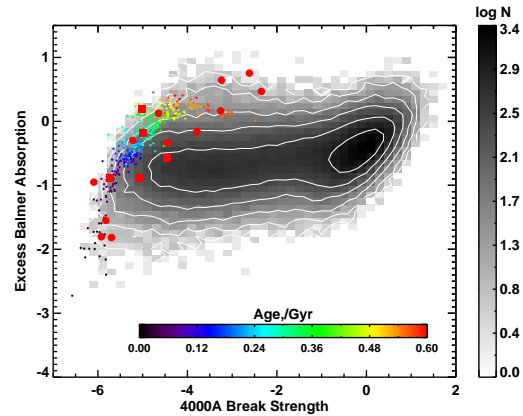} 
\caption{ The plot used to identify the COS-Burst sample from SDSS spectra. The plotted parameters are based on a PCA-based analysis, and represent the strength of the 4000\AA\ break (PC1, on the x-axis) and the excess strength of the high-order Balmer absorption-lines (PC2, on the y-axis), both in dimensionless units. The grey-scale indicates the relative numbers of all SDSS galaxies with the luminosity-weighted mean age of their stellar population increasing from left to right.  The small color-coded dots show a model library of starbursts and post-starbursts with typical burst-mass fractions of 10 to 20\%, durations (e-folding times) of a few hundred million years, and ages increasing from bottom to top. Our COS-Burst sample members are shown as large red symbols.}
\end{figure}

\begin{figure*}
\includegraphics[trim = 10mm 30mm 10mm 20mm, clip,scale=.85,angle=-0]{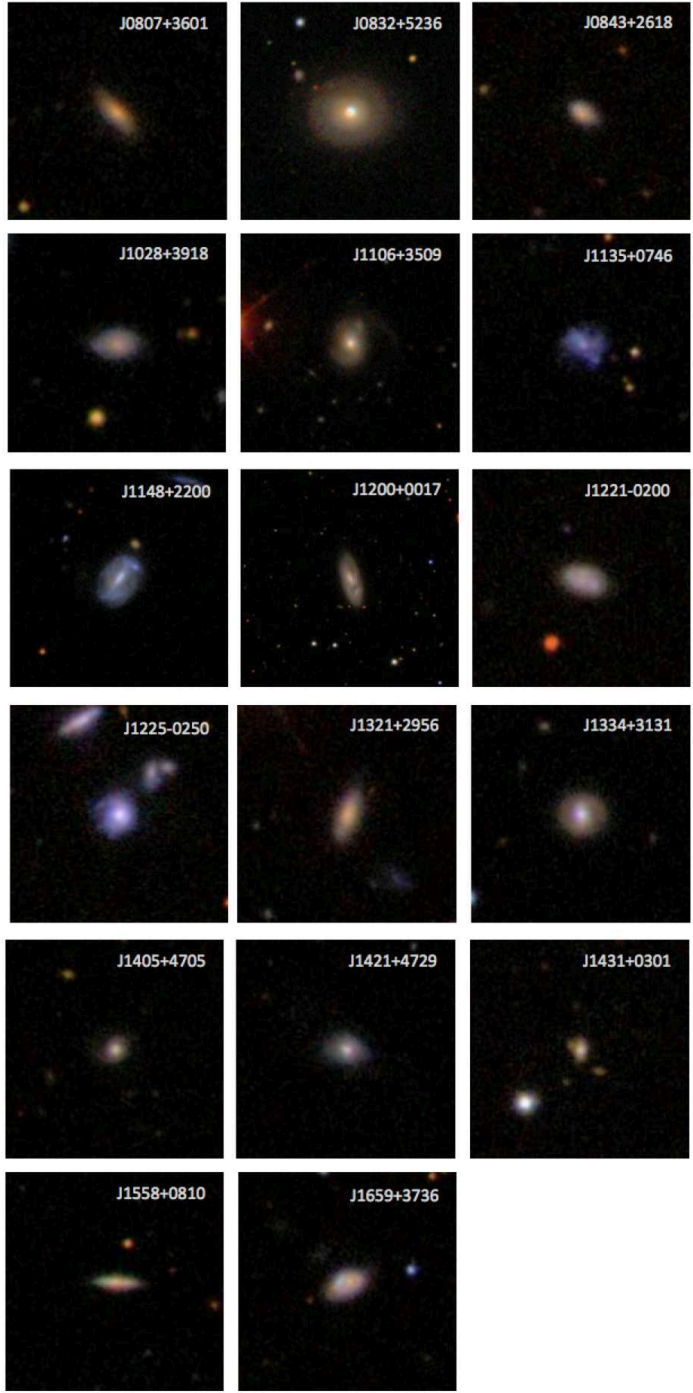}
\caption{Multicolor SDSS composite image of our sample of 17 COS-Burst galaxies. The galaxies are identified by their name on the top right corner. Details of our sample are presented in Table~1. }\label{fig-sample_sdss}
\end{figure*}

\begin{figure}
\includegraphics[trim = 0mm 0mm 0mm 0mm, clip,scale=0.55,angle=-0]{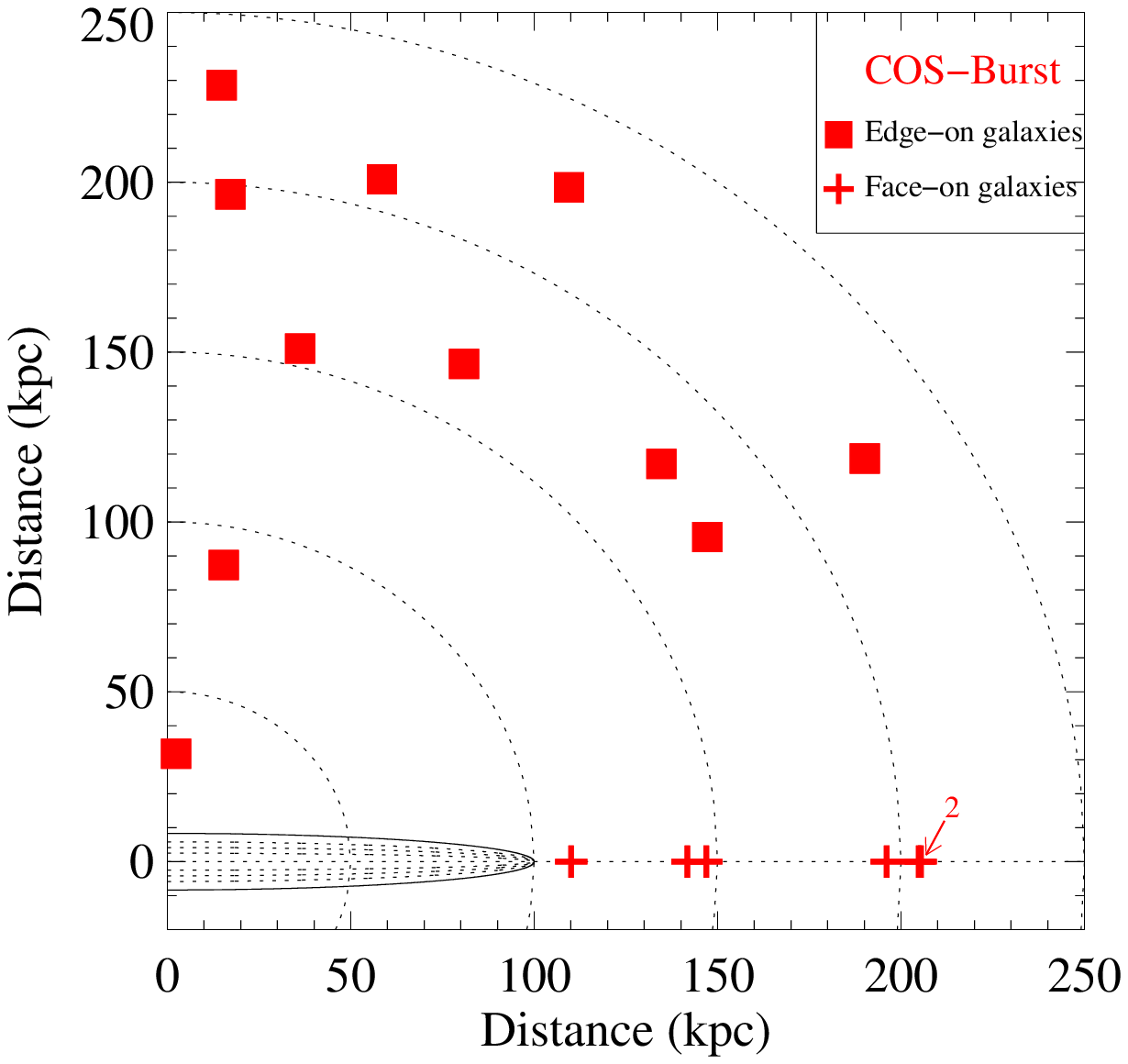}
\caption{ The distribution of the COS-Burst sightlines through the CGM. The orientation of the major axis of the galaxy is shown at the origin (not to scale). In six cases the galaxy is seen close to face-on and these sightlines are plotted as crosses with $y = 0$. We have good coverage of the outer CGM. We do not have sightlines near the galaxy disk plane for the 11 inclined galaxies.}
\end{figure}

\section{Observations}

\subsection{Sample Selection}

We have developed a technique based on Principal Component Analysis (PCA) for identifying starbursts and post-starbursts galaxies using SDSS DR7 spectra (Wild et al. 2007; Wild, Charlot, \& Heckman 2010). The amplitude of the first principal component (PC1) essentially measures the strength of the 4000~$\rm \AA$ break (a probe of the specific SFR over the past few Gyr). The amplitude of the second principal component (PC2) measures the strength/weakness of the high-order Balmer absorption-lines relative to those in an average galaxy with the same value of PC1.  Combining these two gives us a diagnostic diagram for selecting and characterizing galaxies that are undergoing or have recently undergone a strong burst of star formation (typically involving at least 10\% of the galaxy stellar mass).  This is shown in Figure 1, 
 where we also show the time-dependent trajectory of models of such starbursts and their descendants (post-starbursts) over a period of about 600 Myr. A montage of SDSS images of the COS-Burst sample is shown in Figure 2, showing them to mostly be normal late-type galaxies in terms of morphology.

Having selected a sample of SDSS galaxies lying along this trajectory, we have then cross-matched this sample with the SDSS sample of QSOs in the GALEX GR6 catalog. Our final sample consists of those SB/PSB galaxies having a background QSO with FUV $<$ 19 lying along a sightline passing less than 230 kpc \footnote{We adopt H$_0 = 70$ km sec$^{-1}$ Mpc$^{-1}$, $\Omega_{\Lambda} = 0.7$, and $\Omega_M = 0.3$.} from the foreground galaxy. The properties of the individual galaxies in this sample are listed in Table 1, while the median values for the sample-as-a-whole are given in Table 2.  

The geometry of the sightlines probed by the selected COS-Burst targets is plotted in Figure 3. For this figure we use a normalized impact parameter $(\rho/R_{vir}$), where $R_{vir}$ is the halo virial radius. We will describe how we determine $R_{vir}$ below. By chance, we have no sightlines located within $\sim30^{\circ}$ of the galaxy major axis and so cannot probe the effect of the starburst on CGM material near the disk plane. We also primarily sample the outer CGM ($\rho > 0.5$ R$_{vir}$).

\begin{deluxetable*}{ccccc ccccc   ccccc ccccc} 
\tabletypesize{\scriptsize}

\tablecaption{Description of Galaxy Properties. \label{tbl-galaxy}}
\tablewidth{0pt}
\tablehead{
\colhead{Galaxy} &\colhead{Short Name} & \colhead{$\rm z^a$}  &\colhead{$\rm log M_*^b$} &   \colhead{$\rm R_{vir}^c$} & \colhead{$\rho^d$} & \colhead{$\Theta^d$} & \colhead{$\rm f_{burst}^e$} & \colhead{$\rm t_{burst}^e$} & \colhead{$\rm log sSFR^f$} & \colhead{$\rm log SFR^g$} & \colhead{$\rm log pm^h$}\\
\colhead{}   &\colhead{}   & \colhead{}  & \colhead{$\rm Log~M_{\odot}$}  &\colhead{$\rm kpc$} & \colhead{$\rm kpc$} & \colhead{degrees} &\colhead{} & \colhead{$\rm Myr$} & \colhead{$\rm Log~yr^{-1}$}  & \colhead{$\rm log M_{\odot} yr^{-1}$} & \colhead{$\rm dex$}}
\startdata
J080702.28+360141.16 & J0807+3601 & 0.08807 & 10.88 & 259 & 224 & 32 & 0.09$-$0.33 & 600 & -9.57 & 0.99$-$1.53 & 0.33\\
J083228.13+523622.38 & J0832+5236 & 0.01694 & 10.32 & 196 & 205 & $-$& 0.14$-$0.40 & 23$-$53 & -8.20 & 1.35$-$2.12 & 0.29\\
J084356.13+261855.36 & J0843+2618 & 0.11282 & 10.55 & 201 & 179 & 41 & 0.11$-$0.25 & 250$-$432 & -9.30 & 0.91$-$1.24 & 0.21\\
J102846.44+391842.99 & J1028+3918 & 0.11352 & 10.50 & 194 & 89 & 80 & 0.05$-$0.10 & 114$-$280 & -9.43 & 0.51$-$1.07 & 0.25\\
J110624.18+350953.28 & J1106+3509 & 0.07277 & 11.02 & 306 & 142 & $-$ & 0.12$-$0.39 & 600 & -9.42 & 1.22$-$1.92 & 0.32\\
J113522.42+074638.50 & J1135+0746 & 0.08346 & 10.01 & 146 & 155 & 77 &  0.22$-$0.45 & 83$-$189 & -8.58 & 0.71$-$1.42 & 0.24\\
J114848.61+220039.72 & J1148+2200 & 0.03441 & 10.41 & 201 & 209 & 74 & 0.13$-$0.31 & 174$-$341 & -9.12 & 0.37$-$0.82 & 0.24\\
J120018.01+001741.93 & J1200+0017 & 0.02066 & 10.08 & 161 & 32 & 86 & 0.11$-$0.36 & 8$-$38  & -8.08 & 0.86$-$2.00 & 0.43\\
J122115.77-020009.61 & J1221-0200 & 0.06247 & 10.15 & 167 & 197 & 85 & 0.06$-$0.17 & 68$-$205 & -9.11 & 0.32$-$1.04 & 0.32\\
J122534.26-025028.99 & J1225-0250 & 0.06731 & 10.13 & 164 & 196 &$-$ & 0.15$-$0.40 & 68$-$189 &  -8.64 & 1.05$-$1.49 & 0.31\\
J132107.49+295615.39 & J1321+2956 & 0.07228 & 10.48 & 198 & 176 & 33 & 0.04$-$0.09 & 280$-$553 & -9.83 & 0.15$-$0.66 & 0.23\\
J133402.71+313126.87 & J1334+3131 & 0.06310 & 10.34 & 191 & 205 & $-$& 0.13 $-$0.39 & 8$-$38 & -8.00 & 1.78$-$2.34 & 0.41\\
J140502.31+470525.95 & J1405+4705 & 0.14515 & 10.43 & 184 & 147 & $-$& 0.16$-$0.37 & 341$-$553 & -9.24 & 0.84$-$1.19 & 0.21\\
J142120.77+472933.06 & J1421+4729 & 0.07080 &10.08 & 154 & 110 & $-$ & 0.11$-$0.25 & 356$-$600 & -9.47 & 0.05$-$0.61 & 0.20\\
J143140.27+030154.08 & J1431+0301 & 0.15268 & 10.48 & 184 & 229 & 86 & 0.18$-$0.44 & 600 & -9.27 & 1.30$-$1.53 & 0.26\\
J155831.71+081046.37 & J1558+0810 & 0.05832 & 10.11 & 162 & 227 & 61 & 0.27$-$0.47 & 600 & -9.19 & 0.81$-$1.21 & 0.22\\
J165941.51+373654.75 & J1659+3736 & 0.06071 & 10.29 & 184 & 167 & 61 & 0.06$-$0.12 & 220$-$417 & -9.59 & 0.10$-$0.70 & 0.20

\enddata
\tablenotetext{a}{The galaxy redshift from SDSS.}
\tablenotetext{b}{The total galaxy stellar mass from the MPA-JHU Value-Added Galaxy Catalog.}
\tablenotetext{c}{The galaxy halo virial radius, based on the stellar mass. See text for details.}
\tablenotetext{d}{The impact parameter ($\rho$) and orientation with respect to the galaxy major axis ($\Theta$) for the sightline through the CGM. No value is given for $\Theta$ in face-on cases.}
\tablenotetext{e}{The burst-mass fraction and burst age based on PCA analyses of the SDSS spectra. See text for details. The quoted range represents the 16 to 84 percentile values for the probability distribution functions.}
\tablenotetext{f}{The log of the effective specific star-formation rate for the burst, defined as $f_{burst}/t_{burst}$.}
\tablenotetext{g}{The log of the star-formation rate. The quoted range represents no aperture correction to the SDSS fiber (minimum value) and $SFR = sSFR~M_*$ (maximum value).}
\tablenotetext{h}{The uncertainties in $\rm log sSFR$ and $\rm log SFR$ due to the uncertainties in $\rm f_{burst}$ and $\rm t_{burst}$.} 
\end{deluxetable*}

\begin{deluxetable*}{ccccccccccc}
\tabletypesize{\scriptsize}
\tablecaption{Median Properties of the Samples}
\tablewidth{0pt}
\tablehead{
\colhead{Sample$^a$} & \colhead{Number} & \colhead{$\rm log M_*$} & \colhead{$\rm v_c^b$} & \colhead{R$_{50}$} &  \colhead{$\rm R_{vir}$}  &\colhead{$\rho$} & \colhead{$\rm f_{burst}$} & \colhead{$\rm t_{burst}$} & \colhead{$\rm log sSFR$} & \colhead{$\rm log SFR$}\\
\colhead{} & \colhead{} & \colhead{$\rm Log~M_{\odot}$} & \colhead{$\rm km~s^{-1}$} &\colhead{($\rm kpc$} & \colhead{$\rm kpc$} & \colhead{$\rm kpc$} & \colhead{} & \colhead{$\rm Myr$} & \colhead{$\rm Log~yr^{-1}$}  & \colhead{$\rm log~M_{\odot} yr^{-1}$}}
\startdata
COS-Burst & 17 & 10.34 & 129 & 3.2 & 184 & 179 & 0.17 & 280 & -9.24 & 1.07\\
Control 1 & 49 & 10.40 & 140 & 4.0 & 200 & 183 & -- & -- & -10.25 & 0.15\\
Control 2 & 43 & 10.58 & 160 & 4.1 & 223 & 189 & -- & -- & -- & -- \\
Control 3 & 54 & 10.45 & 145 & 4.0 & 205 & 100 & -- & -- & -10.05 & 0.40\\
\enddata
\tablenotetext{a}{The Control 1 sample was used to compare the properties of the Ly$\alpha$ and Si III lines. The Control 2 and 3 samples were for C IV and O VI respectively. See text.}
\tablenotetext{b}{The median value of the characteristic circular velocity of the halo based on the halo mass and virial radius.}
\end{deluxetable*}

\subsection{Generation of Control Samples}

In our analysis we will be emphasizing differential measurements in which each property of CGM of the COS-Burst galaxies is compared to a control sample of normal galaxies. As shown by Borthakur et al. (2016; hereafter B16), there are systematic differences in CGM properties between low-redshift star-forming (blue) and quiescent (red) galaxies (and see  M$\acute{e}$nard et al. 2011 and Bordoloi et al. 2011 for galaxies at intermediate redshifts). Since the COS-Burst galaxies would most likely have been star-forming galaxies prior to the starburst, we will assess the effect of the starburst using only star-forming galaxies in the control samples.\footnote{If we used the quiescent red galaxies instead, the differences we find below would be even larger (B16).}  For the same reason, the control samples were selected to cover the same range in stellar mass ($M_*$) as the COS-Burst sample ($\sim 10^{10}$ to 10$^{11}$ M$_{\odot}$, see Tables 1 and 2). Finally, there are systematic radial declines in the equivalent-widths of the CGM absorption-lines (e.g. Prochaska et al. 2011; Liang \& Chen 2014; Bordoloi et al. 2014a; Johnson, Chen, \& Mulchaey 2015; Borthakur et al. 2015; B16). We have therefore selected control samples with distributions of $\rho/R_{vir}$ that are similar those listed in Table 1 and plotted in Figure 3.

In the analysis below we will be focusing on the properties of the following transitions (which represent the strongest and most commonly detected lines in our data): Ly$\alpha$, Si III 1206.5, C IV 1548.2, and O VI 1031.9. Unfortunately, there is no single control sample that can be constructed from the HST/COS archive that can be used for all four transitions. For Ly$\alpha$ and Si III we will use the sample of star-forming galaxies analyzed by B16 (drawn from the COS-Halos and COS-GASS programs). For C IV we will combine the data for the normal star-forming galaxies in B13 with those in the sample presented in the compilation in Liang \& Chen (2014) that lie in the same range of stellar mass and normalized impact parameter as the COS-Burst sample. For O VI, we select the star-forming galaxies from the compilation of Johnson, Chen, \& Mulchaey (2015) spanning the same ranges in stellar mass and normalized impact parameter as the COS-Burst sample.

\begin{figure*}

\includegraphics[trim=0mm 0mm 3mm 0mm,  clip=true,scale=0.5]{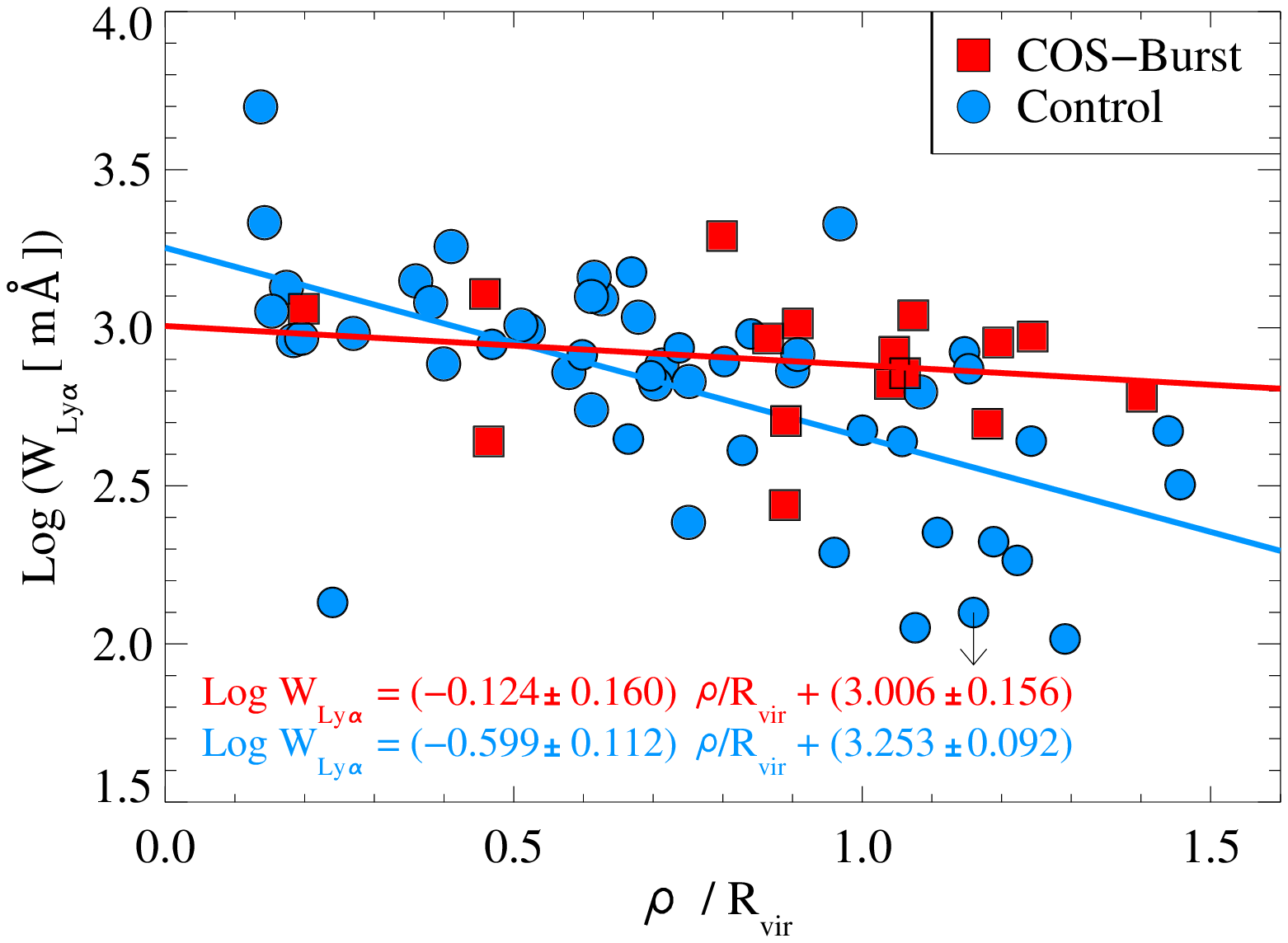}
\includegraphics[trim=0mm 0mm 3mm 0mm,  clip=true,scale=0.5]{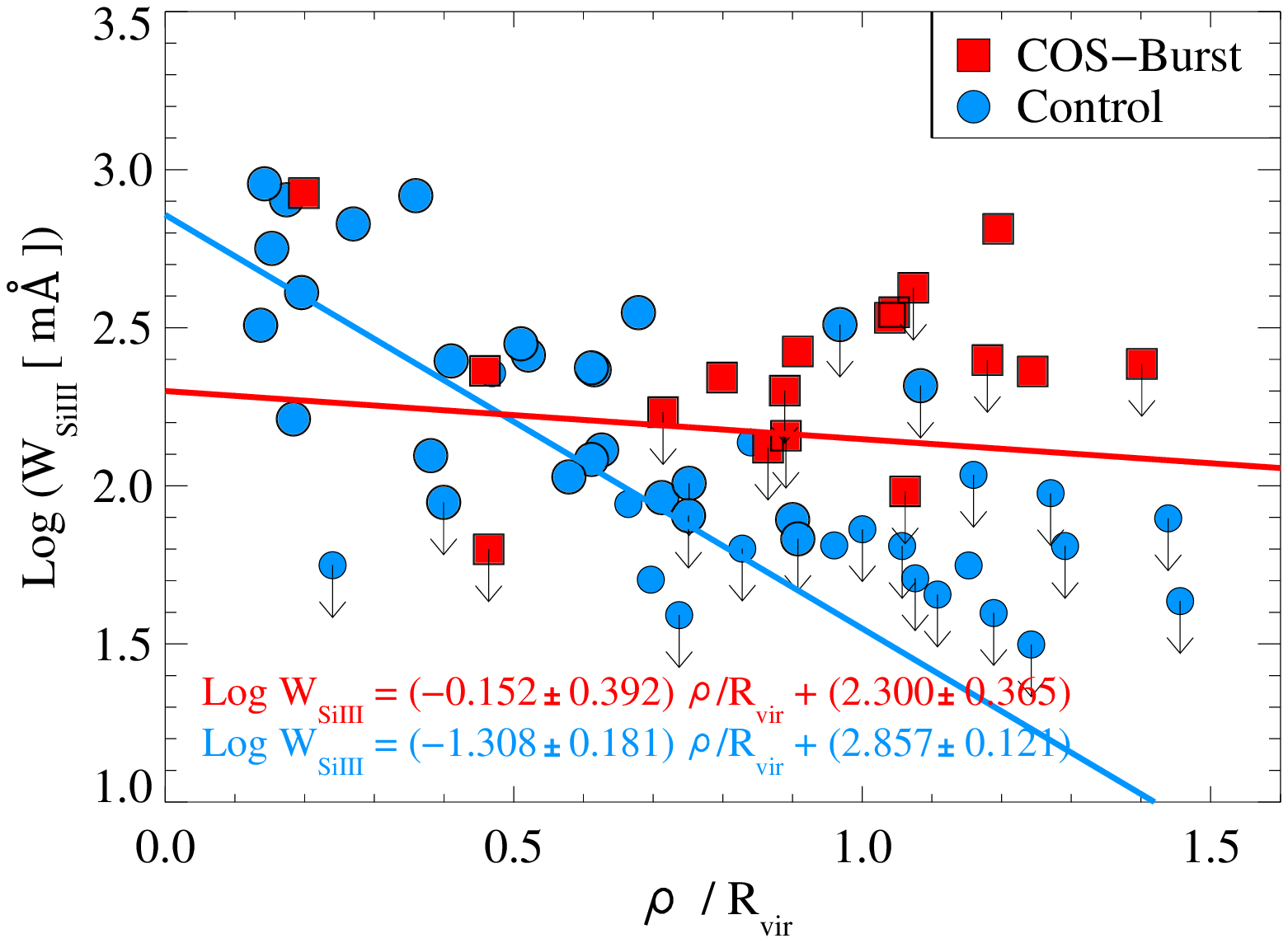}\\

\includegraphics[trim=0mm 0mm 3mm 0mm,  clip=true,scale=0.5]{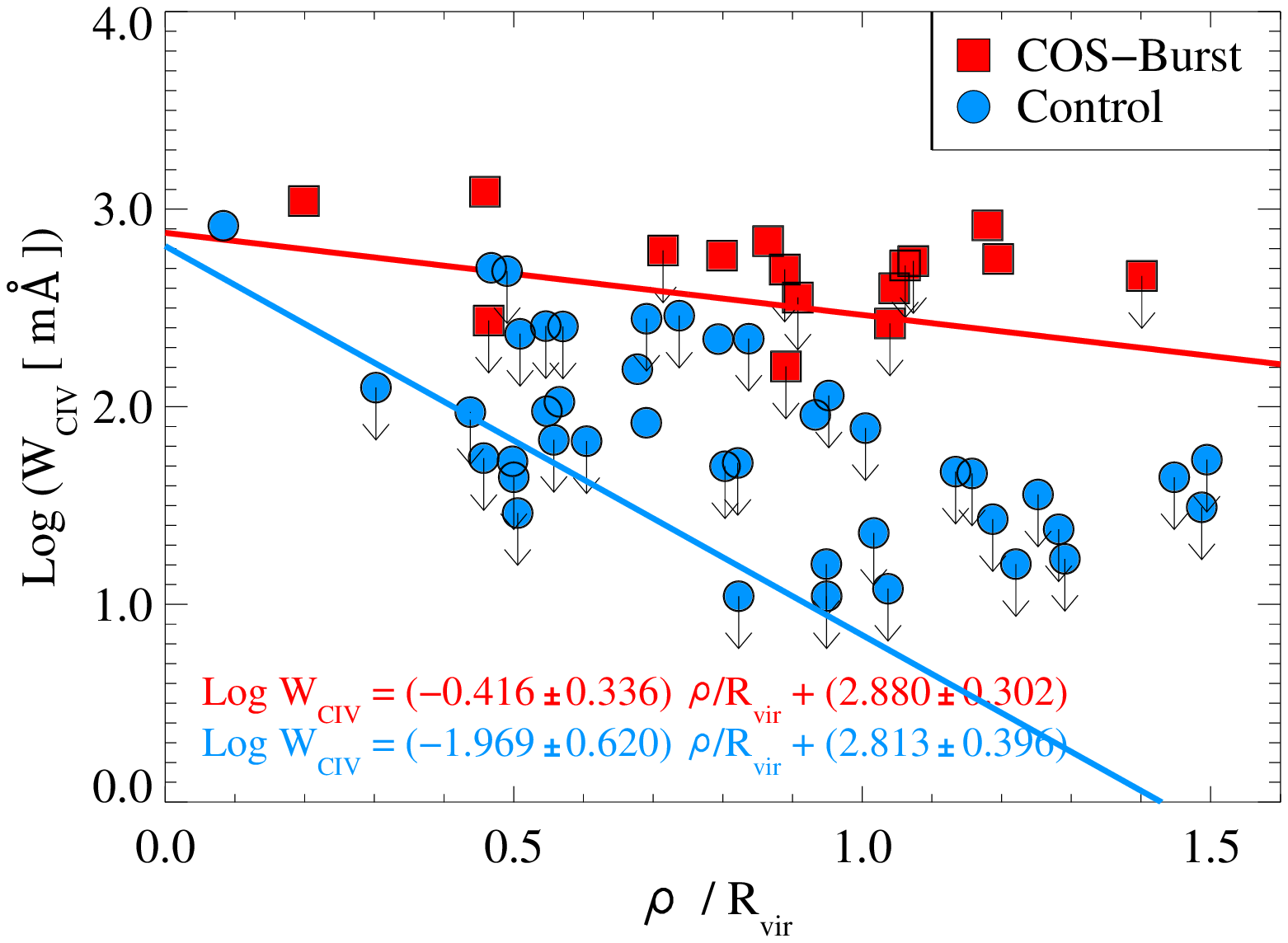}
\includegraphics[trim=0mm 0mm 3mm 0mm,  clip=true,scale=0.5]{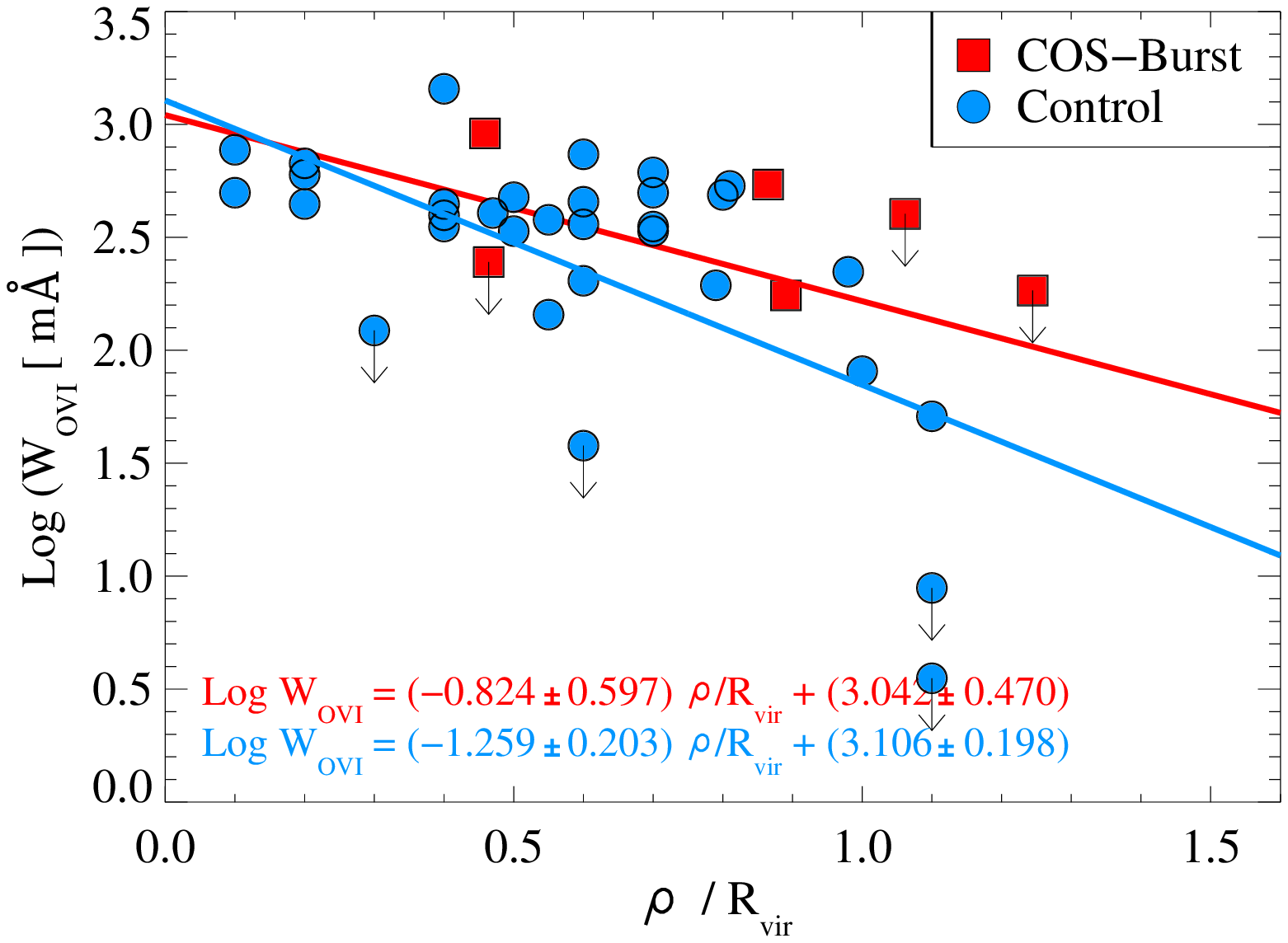}
\caption{We plot the log of the rest-frame equivalent widths of the Ly$\alpha$, Si III 1206.5, C IV 1548.2, and O VI 1031.9 absorption-lines as a function of the normalized impact parameter ($\rho/R_{vir}$) of the sightline through the CGM. The COS-Burst sample of starburst/post-starburst galaxies are plotted as red squares and the control samples of normal star-forming galaxies (see text for details) are plotted as blue circles. In each panel we indicate the best-fit linear relations for both samples (calculated using both detections and upper limits). In all cases, the absorption-lines are stronger in the outer CGM of the COS-Burst galaxies.}
\end{figure*}

The properties of these control samples are compared to those of the COS-Burst sample in Table 2. We will describe how these properties were measured for COS-Burst and the Control samples in section 2.4 below. Table 2 shows that the control samples are good matches to the COS-Burst sample in almost all respects (median values of M$_*$, v$_c$, R$_{vir}$, $\rho$, and galaxy half-light radius ($R_{50}$)). The only exception is the O VI control sample where the median impact parameter is only 56\% as large as for the COS-Burst sample.  

As noted above, none of the COS-Burst sightlines lie within 30$^{\circ}$ of the galaxy major axis. This could potentially bias the comparison to the control samples. For a number of reasons, this should not be a significant effect. First, for the 11 cases in the COS-Burst sample where we can measure the orientation of the COS sightline, we would have only expected 3.7 targets with $\Theta < 30^{\circ}$. 
This only represents 22\% of the sample of 17. Second, over the range of azimuthal angles we do probe ($\Theta = 30^{\circ}$ to 90$^{\circ}$), we see no variation in CGM properties. Third, for the control sample used to compare the properties of the Ly$\alpha$ and Si III lines, Borthakur et al. (2015) showed that there is no azimuthal dependence for the structure of the outer CGM (the region probed in the COS-Burst sample). 
Finally, while an azimuthal dependence of the strength of the Mg II absorption line has been seen in the CGM of star-forming galaxies (Bordoloi et al. 2011; Bouch$\acute{e}$ et al. 2012; Kacprzak et al. 2012; Ho et al. 2016), these sightlines are typically much closer to the disk of the galaxy than in our samples (mean impact parameters of $<$ 50 kpc, 36 kpc, 48 kpc, and 53 kpc for these four studies respectively).

\subsection{Analysis of COS Data}

The new COS Burst data (Program 13862) were obtained using the COS FUV G130M and G160M gratings, yielding spectral resolutions of 15,000 and 18,000 respectively (20 and 18 \kms FWHM). The program was designed so that (when combined with the data in B13) we covered the Si III 1206.5 and \Lya lines in all 17 cases, the C IV 1548.2,1550.8 doublet in 16 cases, and the O VI 1031.9, 1037.6 doublet for the 6 cases with redshifts $z > 0.073$ (placing O VI 1031.9 long-ward of $\sim 1107 \AA$).  

The data were reduced and analyzed following the procedure described in B13, and we refer the reader there for details. We characterized the absorption-line profiles using three non-parametric properties: the equivalent width, the absorbed-flux-weighted mean wavelength (centroid), and the full width at half of the maximum (of the absorbed) intensity. The equivalent widths were then converted to the rest-frame values, the flux-weighted line centroid was used to calculate the velocity difference between the line and the galaxy systemic velocity (based on SDSS spectra), and the line width was converted into \kms. We henceforth refer to these three quantities as $EW$, $\Delta v$, and $FWHM$ respectively. We have also reanalyzed the COS G140M data for the SB/PSB galaxies in B13 to measure these same parameters. The results are listed in Table 3.


\begin{deluxetable*}{ccccccc}
\tabletypesize{\scriptsize}
\tablecaption{Absorption-Line Properties$^a$}
\tablewidth{0pt}
\tablehead{
\colhead{Short Name} & \colhead{$\rm EQ_{Ly\alpha}$}    & \colhead{$\rm EQ_{SiIII}$} & \colhead{$\rm EQ_{CIV}$} & \colhead{$\rm EQ_{OVI}$} & \colhead{$\rm FWHM^b$} & \colhead{$\rm \Delta v^c$}\\
\colhead{} & \colhead{$\rm m\AA$} & \colhead{$\rm m\AA$} & \colhead{$\rm m\AA$} & \colhead{$\rm m\AA$} & \colhead{$\rm km~s^{-1}$} & \colhead{$\rm km~s^{-1}$}}
\startdata
J0807+3601 & $925\pm66$ & $< 132$ & $689\pm127$ & $542\pm124$ & $231\pm16$ & $160\pm75$\\
J0832+5236 & $836\pm36$ & 354$\pm$89 & 398$\pm$ 121 & --- & $212\pm31$ & $54\pm16$\\
J0843+2618 & $507\pm60$ & $<144$ & $<160$ & $175\pm68$ & $116\pm12$ & $93\pm29$\\
J1028+3918 & $1280\pm67$ & $231\pm48$ & $1220\pm72$ & $912\pm122$ & $312\pm31$ & $15\pm24$\\
J1106+3509 & $437\pm23$ & $<63$ & $<273$ & $<246$ & $100\pm20$ & $17\pm56$\\
J1135+0746 & $718\pm61$ & $<96$ & $<519$ & $<402$ & $185\pm21$ & $77\pm25$\\
J1148+2200 & $664\pm123$ & $340\pm120$ & $<264$ & --- & $320\pm20$ & $172\pm22$\\
J1200+0017 & $1148\pm152$ & $843\pm275$ & $1099\pm219$ & --- & $289\pm45$ & $63\pm34$\\
J1221-0200  & $497\pm88$ & $<250$ & $828\pm55$ & --- & $143\pm73$ & $81\pm40$\\
J1225-0250 & $898\pm44$ & $650\pm50$ & $560\pm75$ & --- & $199\pm20$ & $74\pm19$\\
J1321+2956 & $275\pm75$ & $<200$ & $<495$ & --- & $399\pm150$ & $97\pm78$\\
J1334+3131 & $1095\pm23$ & $<423$ & $<546$ & --- & $285\pm10$ & $412\pm64$\\
J1405+4705 & $1950\pm50$ & $220\pm60$ & $582\pm69$ & --- & $412\pm30$ & $25\pm85$\\
J1421+4729 & --- & $<171$ & $<618$ & --- & --- & ---\\
J1431+0301 & $938\pm226$ & $230\pm100$ & --- & $<184$ & $431\pm40$ & $39\pm42$\\
J1558+0810 & $605\pm109$ & $<243$ & $<459$ & --- & $154\pm10$ & $255\pm27$\\
J1659+3736 & $1032\pm269$ & $288\pm154$ & $<357$ & --- & $270\pm25$ & $99\pm18$ 
\enddata
\tablenotetext{a} {These are rest-fram equivalent width for (respectively) the Ly$\alpha$, Si III 1206.5, C IV 1548.2, and O VI 1031.9 lines.}
\tablenotetext{b}{The non-parametric full width at half maximum of the Ly$\alpha$ absorption feature.}
\tablenotetext{c}{The absolute value of the difference between the (absorbed) flux-weighted mean velocity of the Ly$\alpha$ absorption feature and the galaxy systemic velocity from SDSS spectra. The quoted uncertainties include those in both the Ly$\alpha$ centroid and the SDSS galaxy systemic velocity.}
\end{deluxetable*}

\subsection{Parameters from Ancillary Data}

In this paper we will use a number of parameters to characterize the COS-Burst and control galaxies and the QSO sightlines through the CGM (as listed in Tables 1 and 2). The galaxy stellar masses ($M_*$) for the COS-Burst and the Ly$\alpha$/Si III control samples were taken from the MPA-JHU value-added catalog, calculated using the method published by Salim et al. (2007). The stellar masses for the galaxies in the C IV and O VI samples were taken from B13, Liang \& Chen (2014), and Johnson, Chen, \& Mulchaey (2015). The B13 masses were determined the same way as for the COS-Burst sample, while the other masses were based on the NASA-Sloan Galaxy Atlas (http://nsatlas.org/), using a similar methodology.

The stellar mass was used to estimate the mass of the dark matter halo, following the methodology described in B16 for star-forming galaxies (which was based on the analyses of Kravtsov et al. 2014, Liang \& Chen 2014, Johnson, Chen, \& Mulchaey 2015, and Mandelbaum et al. 2016). Following Liang \& Chen (2014) and Johnson, Chen, \& Mulchaey (2015), we have used the halo mass to determine the virial radius ($R_{vir}$) using Equation 3 in Liang \& Chen (2014). This ensures that we have determined the virial radius for the COS-Burst sample in the same way as the Control samples 

We have then defined a normalized impact parameter for each sightline ($\rho_n$), defined as the ratio of impact parameter $\rho$ and $R_{vir}$. We also specify the orientation of the sightline with respect the galaxy major axis (as measured based on SDSS images). Here $\Theta = 0 (90)^{\circ}$ corresponds to a sightline along the galaxy major (minor) axis. 

To determine the parameters of the starburst in the COS-Burst sample we use the SDSS spectra and employ the methodology described in Wild, Charlot, \& Heckman (2010). The starbursts are modeled as events with exponentially-declining star-formation rates. Good fits that include also fitting the observed H$\alpha$ emission-lines required time-constants ($\tau$) of $\sim$ 200 to 350 Myr. The Bayesian analysis yields probability distribution functions for the fraction of the galaxy stellar mass involved in the burst ($f_{burst}$) and the time since the burst began ($t_{burst}$).  We use the median values of the distributions and characterize the uncertainties using the 16 and 84 percentiles in the distribution. While the oldest bursts can be identified, the values of the burst age are not well constrained for $t_{burst} >$ 600 Myr. We define a characteristic star-formation rate averaged over the burst as $SFR = f_{burst} M_*/t_{burst}$ and a characteristic specific SFR per unit mass as $sSFR = f_{burst}/t_{burst}$ in units of inverse years. The uncertainties on these derived parameters are typically $\sim \pm0.3$ dex (see Table 1), based on the probability distribution functions for $f_{burst}$ and $t_{burst}$. We will only use these values to determine the typical (median) values for the sample as-a-whole (as listed in Table 2).

\section{Results}

\subsection{Overview}

\begin{deluxetable}{ccccccc}
\tabletypesize{\scriptsize}
\tablecaption{Rest Equivalent Widths from Stacked Spectra$^{a}$}
\tablewidth{0pt}
\tablehead{
\colhead{$\rm Ly\alpha$} & \colhead{$\rm Si III$} & \colhead{$\rm C IV$} & \colhead{$\rm O VI$} & \colhead{$\rm Si II$} & \colhead{$\rm C II$} & \colhead{$\rm Si IV$}}
\startdata
756$\pm$62 & 165$\pm$43 & 326$\pm$80 & 300$\pm$100 & $<137$ & $<56$ & $<109$\\
\enddata
\tablenotetext{a}{The specific transitions are Ly$\alpha$, Si III 1206.5, C IV 1548.2, O VI 1031.9, Si II 1260.4, C II 1334.5, and Si IV 1393.8. The values are all in $m\AA$.}

\end{deluxetable}

Among the transitions lying in our spectral coverage, we detect the O VI 1031.9, 1037.6, and C IV 1548.2, 1550.8 doublets, and the Si III 1206.5 and Ly$\alpha$ lines.  We have only upper limits for Si II 1260.4, C II 1334.5, and Si IV 1393.8. For these non-detections we have used signal-to-noise-weighted stacked spectra to set upper limits (Table 4). 

For the four detected features, the detection fraction varies. The Ly$\alpha$ line is detected in 16 of the 17 SB/PBS sightlines with contamination by the OI telluric airglow line in J1421+47.  This implies an effective detection-fraction of $f_{det} = $100\%. The Si III line is detected in 8 of the 17 sightlines ($f_{det} =$ 47\% ). The C IV line is detected in 7 of 16 sightlines ($f_{det} =$  44\%). We only probe O VI along 6 sightlines, with 3 detections ($f_{det} =$ 50\%). We note that detections of the different metal lines are strongly related to one another. Five of the 7 Si III detections are detected in C IV, while none of the 9 Si III non-detections are. Similarly, 2 of the 3 O VI detections have detections in C IV, while none of the 3 non-detections do. 

The Ly$\alpha$ lines are optically-thick (saturated), so we do not measure H I column densities. For the detected Si III 1206.5, C IV 1548.2, and O VI 1031.9 lines the median (mean) optical depths in the line cores derived from fitting Voigt profiles are $\sim$ 0.9 (1.1). These optical depths are consistent with the median (mean) value of the C IV 1548.2/1550.8 and O IV 1031.9/1037.6 doublet ratios of 1.73 (1.68), which imply $\tau \sim$ 1 for the stronger member of the doublet in both cases.

Considering only the detected systems, the mean column densities are N$_{O VI} = 10^{14.8}$,  N$_{C IV} = 10^{14.7}$, and N$_{Si III} = 10^{13.5}$ cm$^{-2}$. For species that were not detected, the upper limits based on the stacked spectra (see Table 4) imply column densities of Si II, C II, and Si IV are $<10^{12.9}$, $<10^{13.4}$, and $<10^{13.1}$ cm$^{-2}$, respectively.

\subsection{Starbursts {\it} vs. Controls}

\begin{figure*}
\includegraphics[trim=0mm 0mm 3mm 0mm,  clip=true,scale=0.51]{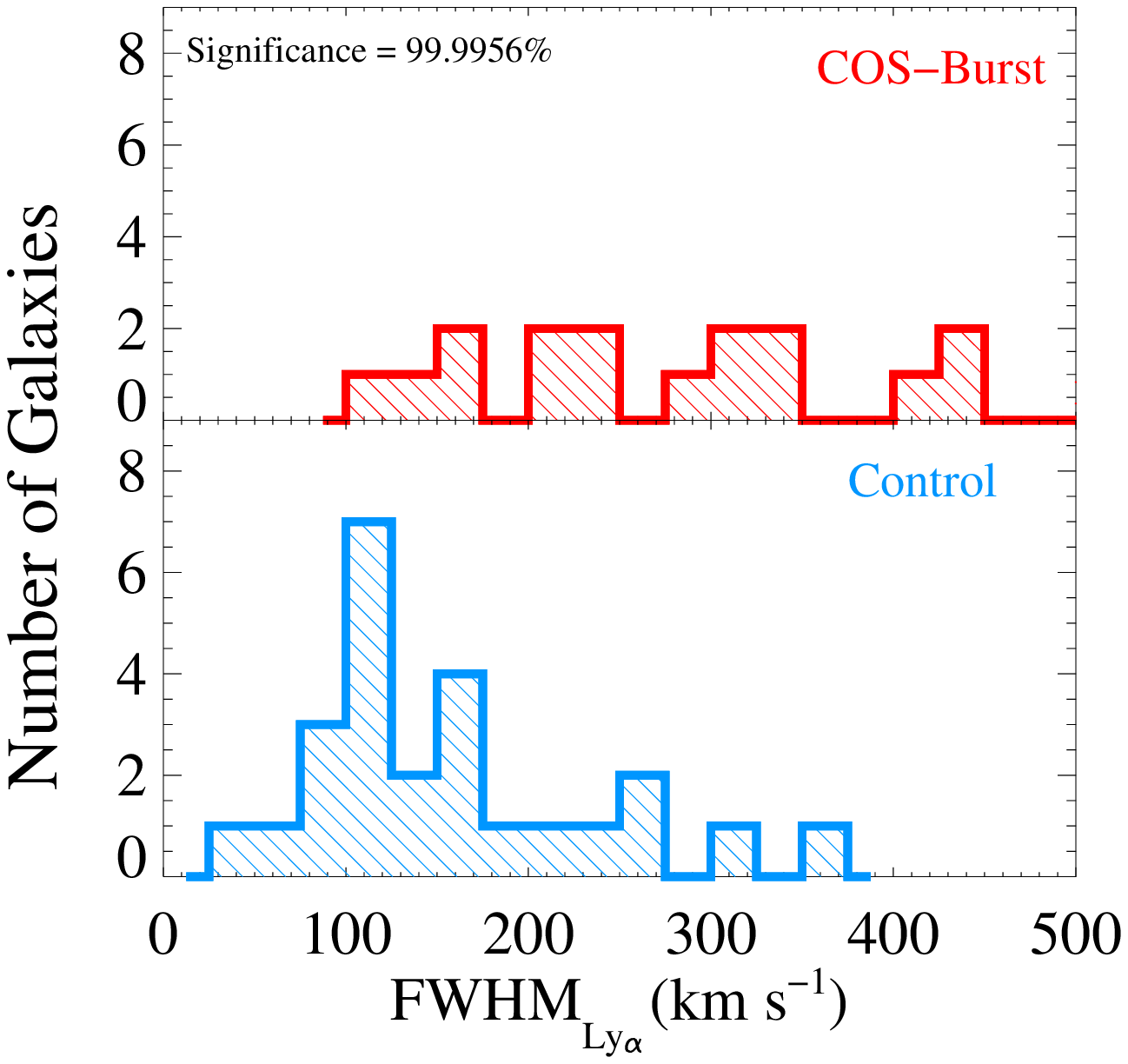} 
\includegraphics[trim=0mm 0mm 3mm 0mm,  clip=true,scale=0.51]{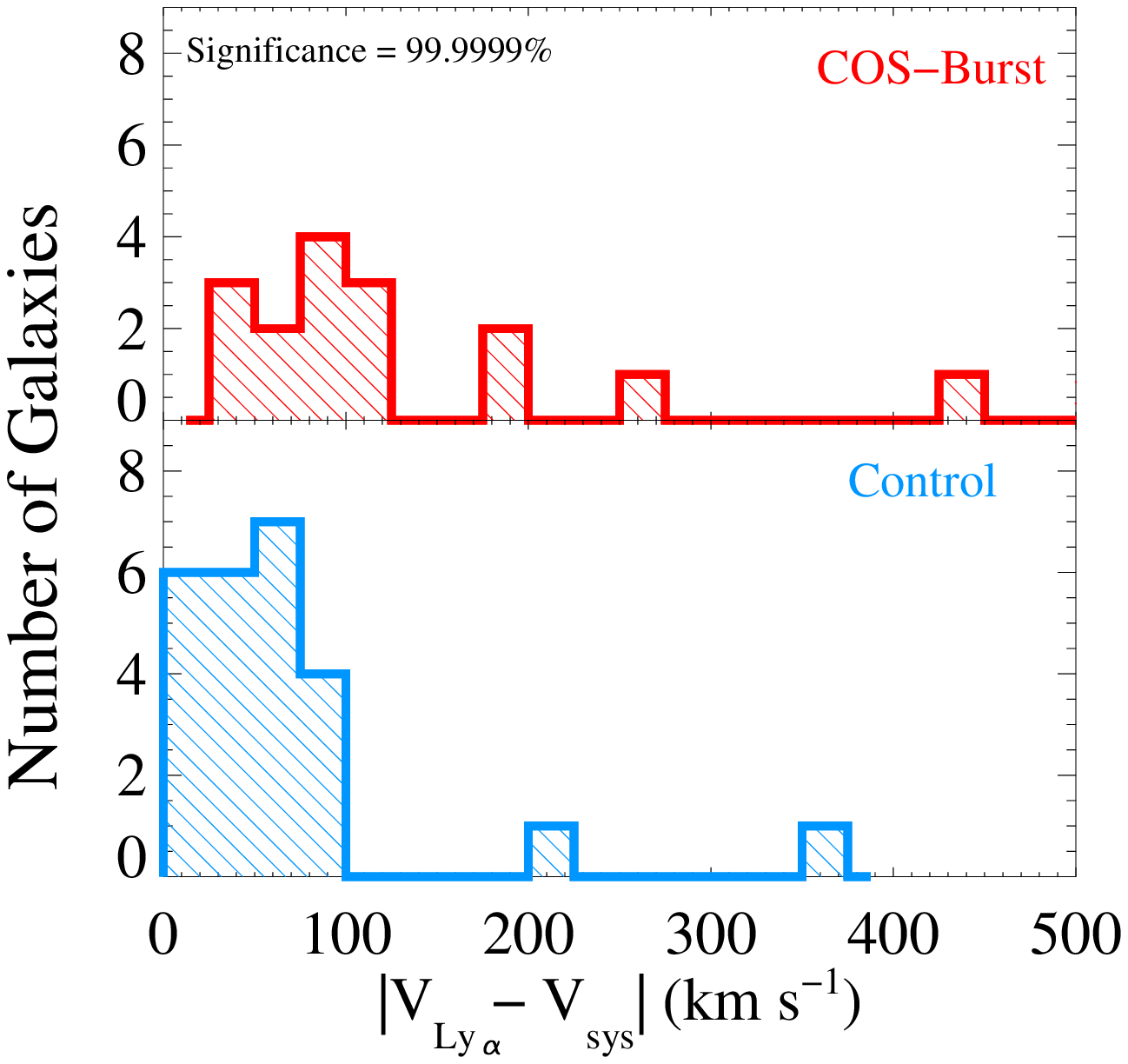}  
\caption{ Left: Histogram of the full-width at half maximum of the Ly$\alpha $ CGM absorption-lines for the COS-Burst galaxies in red and a control sample of normal star-forming galaxies in blue (see text). The lines are broader in the COS-Burst sample at $>$ 99.99\% confidence. Right: Same as left panel, but for the velocity offsets of the Ly$\alpha$ line and the galaxy systemic velocity. The offsets are larger in the COS-Burst sample at the 99.99\% confidence level.}
\end{figure*}

\subsubsection{Radial Distributions}

In Figure 4 we plot the radial distributions of the rest-frame equivalent widths ($EW$) of the Ly$\alpha$, Si III 1206.5, C IV 1548.2, and O VI 1031.9 lines as a function of normalized impact parameter ($\rho/R_{vir}$). In each case, we compare the COS-Burst sample to the relevant control sample described above.  We over-plot the best-fit to the radial dependence of the equivalent width as determined for both the COS-Burst and star-forming (control) galaxies. All these fits make explicit use of the upper limits (see B16 for details). Following B16, we define the excess Ly$\alpha$ equivalent widths as the difference between the logarithms of the measured equivalent width and of the equivalent width at the corresponding normalized impact parameter based on the fit to the control sample.

It is immediately clear that the COS-Burst sample is systematically displaced towards stronger absorption-lines than the control samples in the cases of Ly$\alpha$, Si III, and C IV. The results for O VI are consistent with this, but the small sample size limits the statistical significance. These results pertain mainly to the outer CGM due to the paucity of COS-Burst sightlines in the inner CGM.

\subsubsection{Kinematics}

Since it has the largest number of detections, we have used the Ly$\alpha$ line to characterize the kinematics of the CGM. We do so using two quantities defined above: 1) $\Delta v = |v_{Ly\alpha} -  v_{sys}|$ (where $v_{sys}$ is the systemic velocity of the galaxy based on SDSS), and 2) $FWHM$ defined as the nonparametric full width at half maximum of the absorption-line profile. We have measured these parameters for the COS-Burst sample and for the control sample of star-forming galaxies in COS-GASS (see Borthakur et al. 2015).

In Figure 5 we plot histograms of $\Delta v$ and $FWHM$ for the COS-Burst and control samples. The COS-Burst sample is offset to higher values than the control sample (by median values of $\sim$ 0.3 dex in $FWHM$ and $\sim$0.2 dex in $\Delta v$). These differences are significant at $>99.99\%$ confidence levels according to a Wilcoxon rank test.

To gain more insight into the kinematic properties of the CGM in the COS-Burst sample, we have measured the FWHM of the Ly$\alpha$, Si III, C IV, and O VI absorption lines using the signal-to-noise-weighted stacked spectra for each transition. These stacks were created by aligning the individual spectra using the systemic velocity of the galaxy from SDSS. 
These stacked spectra then show the velocity range covered by the absorbing gas in the entire sample (including both the bulk offsets in velocity and the line-of-sight velocity spreads seen in the individual spectra). The resulting Ly$\alpha$ profile is shown in Figure 5. The FWHM of this line is 424 $\pm$ 20 \kms, and this is consistent with the average value for the noisier Si III, C IV, and O VI stacked profiles (366 $\pm$ 45 \kms). The corresponding stacked Ly$\alpha$ profile for the COS-GASS plus COS-Halos control sample (Table 2) is much narrower (210 $\pm$ 30 \kms).

To put these velocities into context can compare the line widths to expectations for profiles produced by a population of clouds moving randomly through the CGM at the circular virial velocity ($v_c$) of the dark matter halo. For the COS-Burst sample the median value of $v_c$ is 129 \kms (Table 2). In this case, and assuming the halo potential is an isothermal sphere (Binney \& Tremaine 1987), the implied FWHM of the line profile would be 214 \kms. This is only about half as wide as the observed Ly$\alpha$ profile, implying that the observed velocities are super-virial. In contrast, for the control sample, the median value for $v_c$ (140 \kms) would imply FWHM = 233 \kms, consistent with the observed profile.

\subsubsection{Line Ratios}

From Figure 4, it is clear that the differences between the CGM in the COS-Burst galaxies and the normal galaxies are stronger in the Si III and C IV equivalent widths than in Ly$\alpha$. This is shown more explicitly in Figure 7, where we compare histograms of the ratios of the Si III/Ly$\alpha$ and C IV/Ly$\alpha$ equivalent widths between the two samples. 

These differences are consistent with our stacking results and those of Liang \& Chen (2014) for normal star-forming galaxies. For our stacked COS-Burst spectra (Table 4) we find $log(EQ_{CIV}/EQ_{Ly\alpha})=-0.37\pm0.06$ and $log (EQ_{SiIII}/EQ_{Ly\alpha})=-0.66\pm0.11$. The corresponding values from Liang \& Chen (for the radial bin $\rho = 0.56$ to 1.09 $R_{vir}$) are $-0.78\pm0.20$ and $-1.08\pm0.23$ (smaller than the COS-Burst values by $\sim$0.4 dex).   

The physical meaning of the larger ratios of the Si III/L$\alpha$ and C IV/Ly$\alpha$ equivalent widths in the CGM of the COS-Burst galaxies is not straightforward. The Ly$\alpha$ absorption-lines in all the COS-Burst sightlines and the majority of the control sample sightlines are saturated (highly optically-thick). In these cases, the Ly$\alpha$ equivalent width is primarily tracing the spread in velocity of the absorbing gas along the line-of-sight (rather than column density). In contrast, the Si III and C IV lines have typical optical depths of about 1 (see above). Their equivalent widths therefore trace the ionic column densities in the CGM. The enhanced ratios in the COS-Burst sample presumably reflect higher overall gas column densities through the CGM (which increase the strength of the unsaturated lines (C IV and Si III) relative to saturated lines (Ly$\alpha$).

\subsection{The Metal Content of the CGM}

In the following we will estimate the mass of various metal ions using the absorption-line data and the measured column densities. For simplicity, we will simply take the average column density for a given species based on the detections and then multiply this by the fraction of sightlines along which detections were made. We will then multiply this ‘effective column density’ by the geometrical cross-sectional area of the outer CGM to get an implied mass. The median virial radius in our sample was 184 kpc, and we will calculate our masses using an annulus with inner and outer radii of 50 and 200 kpc.

We begin by considering Silicon since we span a wide range of ionization states. The effective column densities are $< 8 \times 10^{12}, 1.5 \times 10^{13}$, and $< 1.2 \times 10^{13} cm^{-2}$ for Si II, Si III, and Si IV respectively, The total implied Silicon mass is M$_{Si} = 4.0$ to 10.0 $\times 10^5 M_{\odot}$ (assuming negligible Si V or higher ions). For Carbon we have effective column densities of $< 2.5 \times 10^{13}$ and 2.2 $\times 10^{14}$ cm$^{-2}$ for C II and C IV respectively, The implied mass is $M_{C IV} = 2.4 \times 10^6$ M$_{\odot}$. Finally, we have an effective column density of 3.2 $\times 10^{14}$ cm$^{-2}$ for O VI, with an implied mass M$_{OVI} = 4.9 \times 10^6~ M_{\odot}$.

These masses can be compared to the metal mass of the warm CGM in normal star-forming galaxies determined by Peeples et al. (2013) for the COS-Halos sample (and see Bordoloi et al. 2014a). Peeples et al. find that the warm phase of the CGM (e.g. as traced by Si II, III, IV) contains a total mass in all metals of $\sim 1.8 \times 10^{-3}$ M$_*$. For the median value M$_* = 2.2 \times 10^{10}$ M$_{\odot}$ for our sample, the implied metal mass in the warm CGM would be $3.9 \times 10^7$ M$_{\odot}$. Adopting a solar value for the Si to metal mass of 0.05 yields $M_{Si} = 1.9 \times 10^6$ M$_{\odot}$, somewhat larger than we estimate in the outer CGM in our sample. The difference (as seen in Figure 4) is the radial distributions: the metals assayed by Peeples et al. in normal star-forming galaxies are largely confined to the inner CGM (interior to $\sim 0.5 R_{vir}$), whereas the metals detected in the CGM of the COS-Burst sample extend out to beyond the virial radius. We will discuss the implications of this below.

Estimating the total gas mass is more uncertain since it depends upon assumptions about the metallicity and on ionization corrections. Based on Silicon, we estimate that the total mass of the gas traced by Si II, III, and IV in the outer CGM is M$_{tot} = 0.57$  to 1.2 $\times 10^9$ M$_{\odot}$  Z$_{\odot}/$Z, where Z is the gas-phase Silicon abundance in the CGM and Z$_{\odot}$ is its  solar value.  Werk et al (2014) find a median metallicity of 0.2 solar in the CGM of COS-Halos galaxies, which implies M$_{tot} \sim 2.8 $ to 6.3 $\times 10^{9}$ M$_{\odot}$ for the outer CGM in the COS-Burst sample.  This is a bit smaller than the mass estimated by Werk et al. for the inner CGM in typical galaxies.

\section{Discussion}

\subsection{Introduction}

We have found clear evidence that the outer CGM around the starbursts differs systematically compared to the control galaxies. The Ly$\alpha$, Si III, C IV, and possibly O VI absorption-lines are stronger as a function of normalized impact parameter, and the ratios of the equivalent widths of Si III/Ly$\alpha$ and C IV/Ly$\alpha$ are larger than in normal star-forming galaxies of the same stellar mass. Both the widths and the velocity offsets (relative to $v_{sys}$) of the absorption-lines are also significantly larger in the starbursts than in the control galaxies. In fact, the implied velocities in the CGM of the COS-Burst galaxies are roughly twice the halo virial velocity, implying that some force other than gravity is affecting the dynamics.

It is important to recognize that the physical meaning of the line strength is not the same for all the lines. As noted above, the Ly$\alpha$ absorption-lines are generally saturated, so the Ly$\alpha$ equivalent width is primarily tracing the spread in velocity of the absorbing gas along the line-of-sight (rather than column density). Thus, the stronger Ly$\alpha$ lines are connected to the different kinematic properties of the CGM in the COS-Burst galaxies noted above. In contrast, the Si III, C IV, and O VI lines have roughly unit optical depth. Their equivalent widths therefore trace the ionic column densities and their detection fraction probes the covering factor of this gas in the CGM. The results above therefore mean that there are both more ionized metals and larger characteristic velocities in the outer CGM of the COS-Burst galaxies compared to the controls.

In the sections below, we will consider various mechanisms that could link the starburst to the properties of the CGM.

\subsection{Alternatives to a Wind Model}

In the next section, we will explore in some detail a model in which the CGM in our sample of COS-Burst galaxies is being affected by a galactic wind driven by the starburst. Before doing so, we want to consider possible alternative interpretations. Firstly, we have shown that the CGM around the COS-Burst galaxies has unusual properties. But what is the direction of the causal connection: is the starburst producing an unusual CGM or is an unusual CGM fueling the starburst?

One possibility is that these starbursts have been triggered by major mergers that have affected the CGM that now surrounds the merger. This is not plausible in the case of the COS-Burst sample, for a number of reasons. First, this idea would not explain the super-virial velocities we observe. Moreover, as can be seen in the imaging montage in Figure 2, the members of the COS-Burst sample are mostly normal late-type galaxies. A few appear to be interacting with companions, but few (if any) appear to be recent or on-going mergers. This is consistent with the fact that only about 12\% of starbursts with star-formation rates like those of the COS-Burst sample are triggered by mergers (Sanders \& Mirabel 1996).

Another possibility is that the environment on the COS-Burst galaxies differs systematically from that of normal star-forming galaxies. For example, Johnson et al. (2015) found that there are differences in the CGM between isolated galaxies and galaxies in groups. To evaluate this, we have used the SDSS DR7 group catalog of Tago et al. (2010). This shows that 7 of the 17 (41\%) COS-Burst galaxies lie in groups. We have verified that there are no significant differences in the group {\it vs.} isolated COS-Burst galaxies in Figures 4, 5, and 7 in this paper. Moreover, we find that about 60\% of the Control galaxies from COS-GASS are in groups. 

A more general, and simple, way to consider whether the causal connection is an inward one is to ask whether it is plausible for the inflow or outflow of mass, metals, and energy to occur on the relevant timescale. The median value for the impact parameter in the sample is $\rho =$ 179 kpc and the median starburst age is $t_{burst} \sim $ 280 Myr. A causal connection then requires a characteristic velocity of $\rho/t_{burst} >$ 630 km sec$^{-1}$. This represents the minimum velocity required for the outer part of the CGM to be causally connected to the starburst. This is considerably larger than the median halo circular velocities (129 km sec$^{-1}$). This makes it implausible that the causal connection is one in which the unusual properties of the outer CGM are related to the fueling of the starburst (e.g. an inward flow driven by gravity). 

Of course, the innermost part of the CGM could be causally connected via inflow to the starburst. Observations of the inner CGM ($\rho <$ 50 kpc) of star-forming galaxies at intermediate redshifts show enhanced Mg II absorption compared to quiescent galaxies, and the Mg II absorption is enhanced along the galaxy minor and major axis (Bordoloi et al. 2011; Bouch$\acute{e}$ et al. (2012), Kacprzak et al. 2012; Ho et al. 2017). This may imply that a causal connection between the star-formation and the CGM in both directions (outflow along the minor axis and inflow along the minor axis) may be happening in the inner CGM. Unfortunately, we have only one sightline interior to 50 kpc, and none near the galaxy disk plane.  

Returning to the outer CGM, the characteristic velocity required for causal connection can be more easily accommodated if the starburst is affecting this region (reflecting an outward flow). In this case, the flows of energy/mass/metals at speeds well in excess of $v_c$ are possible (as we will show below).

We will now consider the effects of radiation from the starburst, which could very rapidly reach the outer CGM. We will first consider the effects of ionizing radiation on the physical state of the CGM, and then the effects of non-ionizing UV radiation on the CGM dynamics. The CGM in normal galaxies is believed to be photo-ionized by the meta-galactic background (e.g. Werk et al. 2013, 2014). To assess whether the additional ionizing radiation from the starburst could be important, we can compare the relative intensities of these two sources. Following B13, this contribution from the starburst can be written as:

\begin{equation}
\Phi_{SB} = 1.4 \times 10^5~SFR r_{100}^{-2}~f_{esc}~cm^{-2} s^{-1}
\end{equation}

Here, SFR is measured in units of M$_{\odot}~ yr^{-1}$, the distance from the starburst to the CGM is measured in units of 100 kpc, and $f_{esc}$ is the fraction of ionizing photons that escape the starburst and reach the CGM.  Using the median values for these parameters in Table 3 for our COS-Burst sample (SFR = 12 M$_{\odot}$ year$^{-1}$, $r_{100} =$ 1.79), and the value $\Phi_{MGB} = $ 2750 cm $^{-2}$ sec$^{-1}$ from Haardt \& Madau (2012), we find that ratio $\Phi_{SB}/\Phi_{MGB} = 192 f_{esc}$. In {\it typical} starbursts, there are only upper limits on $f_{esc}$ (one to a few \% - e.g. Heckman et al. 2011). 
\footnote{If the material in the outer CGM has been transported outward, 
the intensity of the ionizing radiation from the starburst would have been initially much larger (e.g. a factor of three smaller distance from the starburst would imply a radiation field nearly an order-of-magnitude larger). On the other hand, even if material near the starburst was initially photo-ionized by the starburst, the recombination times for the metal ions are so short that the material would recombine long before it reached the outer CGM. For the CGM cloud densities inferred for the inner CGM of $\rm 10^{-3}~ cm^{-3}$ (Werk et al. 2014), the recombination times for C IV to C III (Si III to Si II) would only be about 4(10) Myr (Nahar 1995; Nahar, Pradhan,\& Zhang 2000).}

While it is therefore possible that the starburst is contributing significantly to the photo-ionization of the CGM (for large enough values of $f_{esc}$), this would not naturally account for the stronger Si III we see in the outer CGM around the COS-Burst galaxies. As we have demonstrated above, the outer CGM in these galaxies contains a substantially larger mass of Si than is present there in normal galaxies. In both the COS-Burst and normal galaxies (Werk et al. 2014), the amount of Si III in the CGM is similar to or greater than the amounts of Si II and Si IV. Simply increasing the intensity of the ionizing radiation field in the CGM (thereby converting Si II to Si III and Si III to Si IV) cannot explain this difference in Si mass. 

The CGM of the COS-Burst galaxies also differs from that of normal galaxies in terms of its kinematics (higher, super-virial velocities).  In principle this could be the result of the acceleration of CGM clouds by the pressure of the far-UV radiation emitted by the starburst (e.g. Murray et al. 2005). Given the typical Si column densities in the CGM estimated above, and assuming a standard dust-to-metals ratio (Mattson et al. 2014), the implied optical depth of the CGM clouds to far-UV radiation would only be of-order 0.002. Assuming an isothermal potential characterized by a velocity $v_c$, the ratio of the force due to radiation pressure to the force of gravity acting on a cloud is given by:

\begin{equation}
F_{rad}/F_{grav} = L \tau/(4 \pi c r  v_c^2 N_H \mu)
\end{equation}

Here $L$ is the UV luminosity of the starburst, $\tau$ is the cloud dust optical depth, $r$ is the distance between the cloud and the starburst, $v_c$ is the halo circular velocity, $N_H$ is the cloud hydrogen column density, and $\mu$ is the mean mass per particle (1.4 $m_H$) in the cloud. We use the parameters representing the medians in our sample ($L \sim 10^{43}$ erg sec$^{-1}$, $\tau$ = 0.002, $r =$179 kpc, $v_c =$ 129 km sec$^{-1}$). The column density of Si estimated above implies a total Hydrogen column density of $N_H = 7 \times 10^{17} (Z_{\odot}/Z$) cm$^{-2}$ (since the lower the metallicity, the higher the implied value of $N_H$ for a given $N_{Si}$). We then find $F_{rad} = 3.5 \times 10^{-4} ~Z/Z_{\odot}~ F_{grav}$. We conclude that radiation pressure will be dynamically negligible.

\subsection{A Galactic Wind}

We now assess the possible mechanisms by which a starburst-driven wind could affect the CGM. The physics of galactic winds driven by a population of massive stars has recently been reviewed by Heckman \& Thompson (2017). In addition to radiation pressure, the kinetic energy and momentum associated with stellar winds and supernova explosions play crucial dynamical roles. The stellar ejecta, through supersonic collisions, will create a hot volume-filling fluid inside the starburst (Chevalier \& Clegg 1985) with a temperature given by $T \sim 4 \times 10^7 {\rm K} (\alpha/\beta)$. Here $\alpha$ is the fraction of the kinetic energy injected by the massive stars that is not lost to radiative cooling and $\beta$ is the ratio of the mass-injection rate to the SFR ($\beta \sim$ 0.3 corresponds to the pure stellar ejecta). 

This hot gas will expand along the minor axis of the galaxy disk and blow out into the halo. The wind fluid will then cool adiabatically as it expands and will reach a terminal velocity given by $v_{term} \sim 1500$ km sec$^{-1} (\alpha/\beta)^{1/2}$.  In the discussion to follow we take $\alpha =$ 1 and $\beta = 0.3$, consistent with detailed modeling of the M 82 wind (Strickland \& Heckman 2009). This implies $v_{term} =$ 2800 km sec$^{-1}$. This velocity is high enough, in principle, for a wind to traverse the CGM on a timescale much shorter than the typical starburst lifetime.

One way to explain the differences between the CGM in the COS-Burst galaxies compared to normal star-forming galaxies is that a significant amount of metals have been transported to the outer CGM. Here we ask whether this is feasible based on the available energy delivered by a galactic wind. To set the stage, we note that for the median star-formation rate and burst ages in our COS-Burst sample the implied kinetic energy released by supernovae and stellar winds will be $\sim 8 \times 10^{58}$ ergs (Heckman \& Thompson 2017).

Above, we have estimated that the total gas mass in the outer CGM of the COS-Burst galaxies is about 5 $\times 10^{9} M_{\odot}$. The work required to move this much mass by a factor of two in radius in an isothermal gravitational potential with $v_c =$ 129 km sec$^{-1}$ would be about $10^{57}$ ergs. This is almost two orders-of-magnitude smaller than the potentially available energy, so in principle a starburst-driven wind could re-arrange the CGM. We now examine this model in more detail.

\begin{figure}
\includegraphics[trim=0mm 0mm 3mm 0mm,  clip=true,scale=0.45]{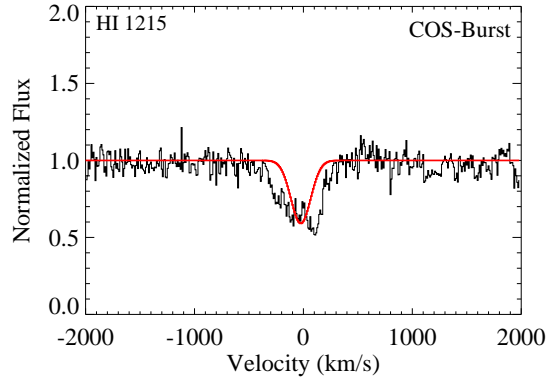} 
\caption{ The stacked Ly$\alpha$ profile for the sample of 17 COS-Burst galaxies. The FWHM of this line is 424$\pm$25 \kms. Over-plotted in red is the profile expected for a population of absorbers moving randomly through the CGM at the median halo virial velocity for the sample. This profile has a FWHM of 214 \kms (only half the observed value). Note that the asymmetric structure of the stacked profile reflects small number statistics in the stack.}
\end{figure}

\subsubsection{A Wind-Cloud Model}

The ram pressure of the wind fluid, combined with radiation pressure, will accelerate gas clouds and drive them outwards (e.g. Murray et al. 2005; Chevalier \& Clegg 1985). This process is believed to be responsible for the blue-shifted interstellar absorption-lines (seen in the “down-the-barrel” observations of starbursts) and the high-velocity optical emission-line gas seen on scales of $\sim$ 1 to 10 kpc along the minor axes of starburst galaxies. These scales are far smaller than those we are probing here. Numerical simulations show that clouds accelerated in this way are unlikely to survive long enough to be transported over such large distances (see the discussion in Heckman \& Thompson 2017), so it is not plausible that clouds launched near the starburst could reach the outer CGM intact.

Instead we will consider a model in which we are observing the impact of the wind fluid on the {\it pre-existing} clouds in the CGM, with properties like those derived by Werk et al (2014). While the origin of these clouds is uncertain, the COS data provide very useful empirical information about their properties. We will assume that the clouds have a relatively small volume filling-factor, and that any diffuse volume-filling phase of the CGM can be ignored. In the next subsection we will relax this assumption. To characterize the wind, we will use parameters appropriate to the median values in our sample (Table 2). 

As the hot wind fluid flows out into the CGM, its ram pressure can accelerate the pre-existing CGM absorption-line clouds it has overtaken.  As in the case of radiation pressure considered above, we can compare the ram pressure and gravitational forces acting on a CGM cloud:

\begin{equation}
F_{ram}/F_{grav} = \dot{p}_{wind}/(4 \pi r v_c^2 N_H \mu)
\end{equation}

Here $\dot{p}_{wind}$ is the momentum flux carried by the wind. For a star-formation rate in M$_{\odot}$/year, this is given by:

\begin{equation}
\dot{p}_{wind} = 1 \times 10^{34}(\alpha \beta)^{1/2}  SFR  ~dynes 
\end{equation}

Using the median values for these parameters in the COS-Burst sample (Table 3), and taking $\alpha = 1$ and $\beta = 0.3$, we find $F_{ram}/F_{grav} \sim 34 (Z/Z_{\odot})$. Unless the metallicity in the CGM clouds is much lower than estimated by Werk et al. (2014) and Prochaska et al. (2017), the wind can overcome gravity and accelerate the clouds outwards. This flow could in principle carry metals outward and account for the higher mass of metals in the outer CGM of the COS-Burst galaxies.

Heckman et al. (2015) derived the equation of motion for a cloud accelerated by the combined inward force of gravity and outward force due to a wind. In their notation, the ratio of the starburst momentum flux to the minimum needed to balance gravity is $R_{crit} = F_{ram}/F_{grav}$. They showed that the maximum velocity to which the cloud could be accelerated is given by:

\begin{equation}
v_{max} = \sqrt{2} v_c [(R_{crit} - 1) - lnR_{crit}]^{1/2}
\end{equation}

\begin{figure*}
\includegraphics[trim=0mm 0mm 3mm 0mm,  clip=true,scale=0.51]{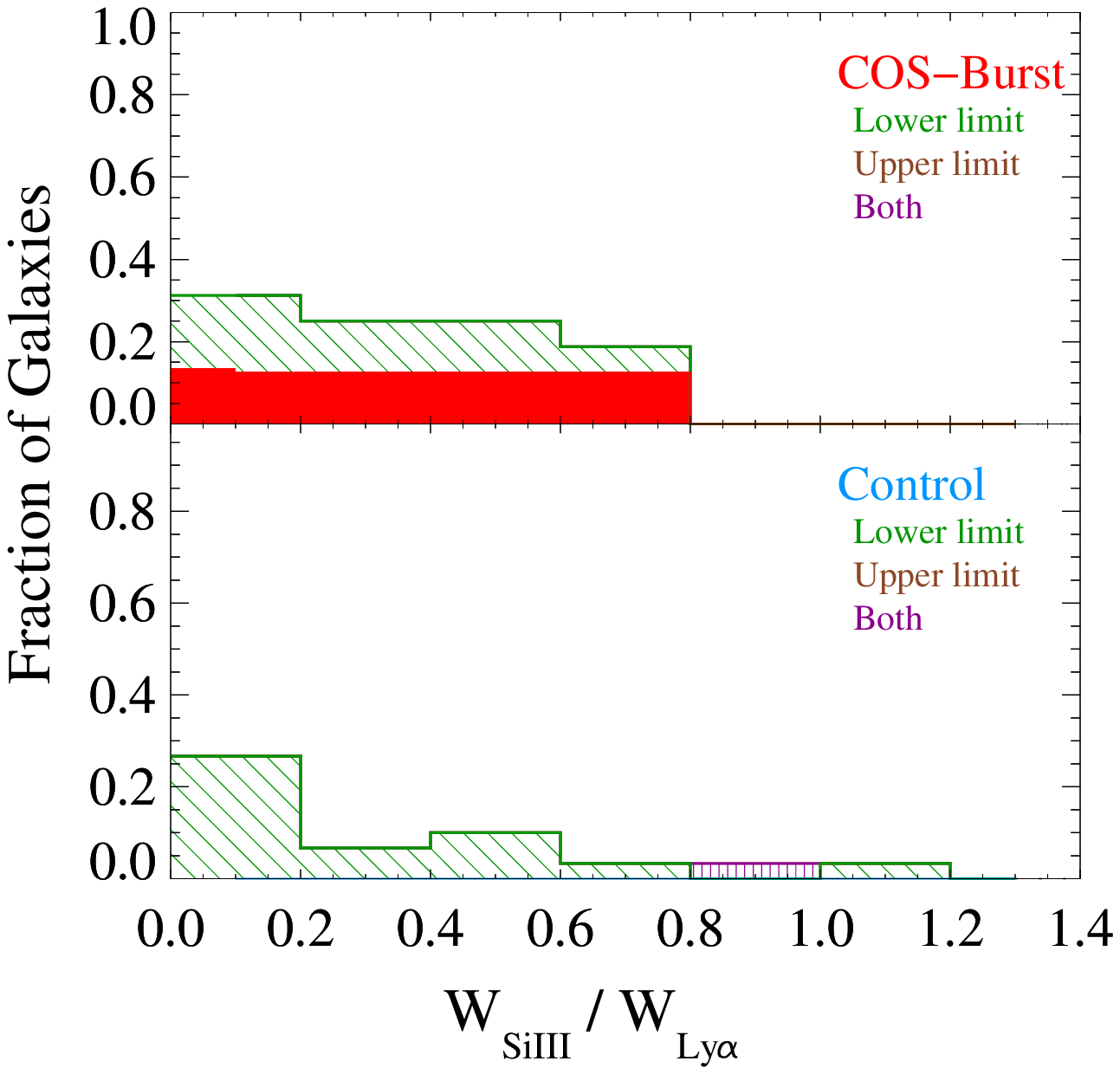} 
\includegraphics[trim=0mm 0mm 3mm 0mm,  clip=true,scale=0.51]{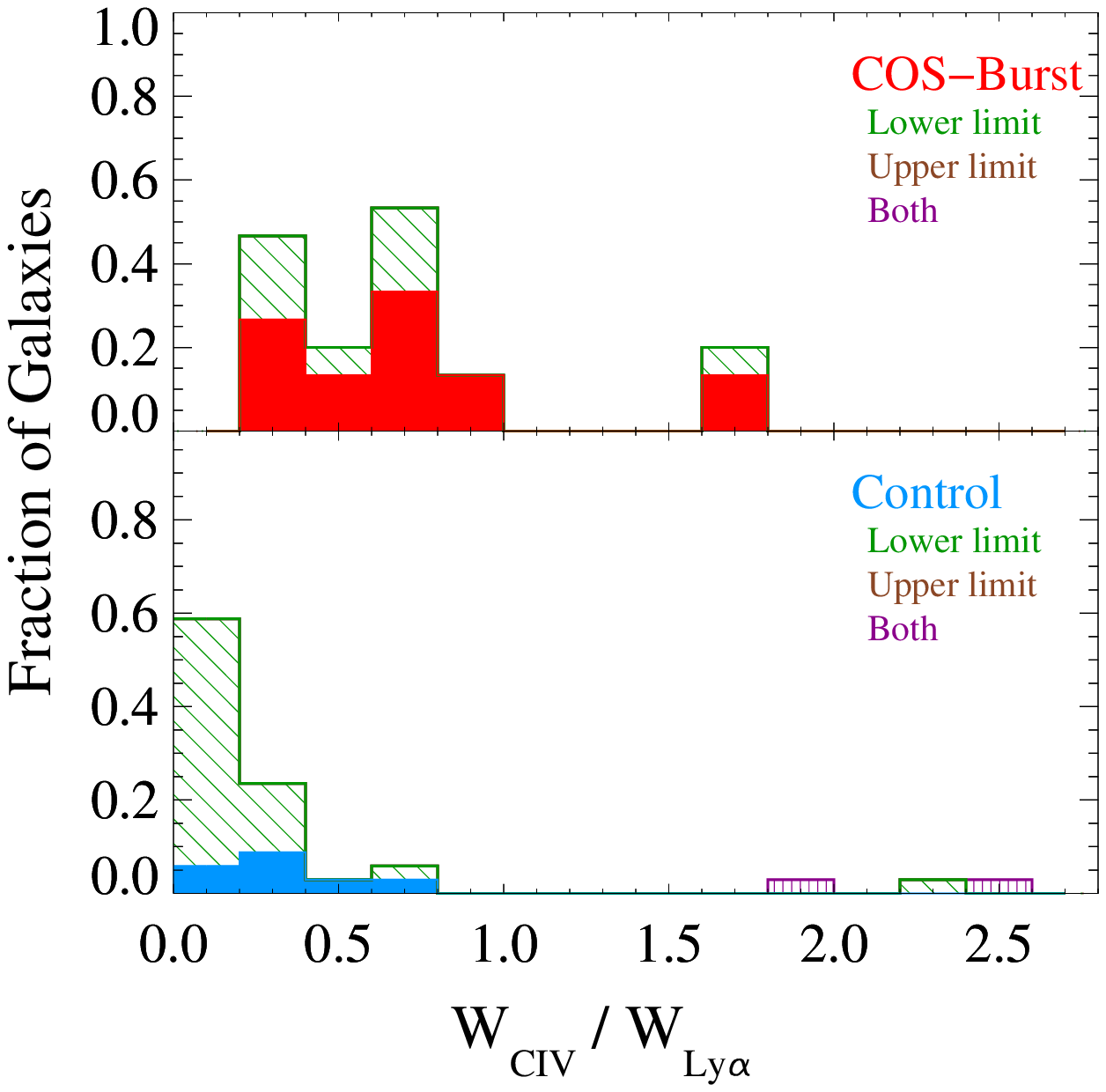} 
\caption{ Histograms of the ratios of the Si III/Ly$\alpha$ and C IV/Ly$\alpha$ equivalent widths for the COS-Burst galaxies (top panels) and control star-forming galaxies (bottom). The cases with detections of the metal lines are shown in solid red (top) and solid blue (bottom). The green hatching show the upper limits when the metal lines were not detected. Both ratios are higher in the COS-Burst sample. A Wilcoxon Rank Test shows that the median values are larger for the COS-Burst sample at the 95\% (Si III) and $>$99.99\% (C IV) confidence level.}
\end{figure*}

If we adopt $Z_{Si} \sim 0.3 Z_{\odot}$ (Prochaska et al. 2017) to evaluate $R_{crit}$ (see above), this predicts a maximum outward velocity of  $\sim 3.7 v_c$, or $\sim$ 480  km sec$^{-1}$ for the median value of $v_c =$ 129 km sec$^{-1}$. For a median starburst lifetime of 280 Myr, a cloud moving at 480 \kms could travel a distance of about 135 kpc ($\sim$ 0.7 R$_{vir}$). Thus, the large-scale transport of CGM clouds by this mechanism is at least potentially feasible.

To compare this outflow velocity to the width of observed stacked Ly$\alpha$ profile (Figure 6) requires a translation of the outflow velocity into a profile of the projected line-of-sight velocities. To do this, we have constructed a simple numerical model in which a spherically-symmetric mass-conserving outflow travels at a constant velocity $v_{out}$, for a time sufficient to reach a maximum radial extent of $R_{max}$. We have then measured the resulting FWHM along a line-of-sight through the outflow as a function of $\rho/R_{max}$. We take $R_{max} \sim 1.5 R_{vir}$ (the maximum value for $\rho$ in our sample). For the median value of $\rho/R_{max}$ in our sample (0.63), the observed FWHM of 424 \kms implies $v_{out} = 352$ \kms. 

The wind will drive shocks into the clouds. Momentum balance across the shock implies that the shock velocity driven into a cloud initially at rest, by a wind flowing at $v_{wind}$, will be given by:

\begin{equation}
v_{cloud,s} = v_{wind}(n_{wind}/n_{cloud})^{1/2}
\end{equation}

Here, $n_{cloud}$ and $n_{wind}$ are the cloud and wind particle densities. Werk et al. (2014) find $n_c \sim 10^{-4}$ cm$^{-3}$ in the outer CGM. Adopting the wind parameters as above, we calculate $n_w \sim 10^{-7}$ cm$^{-3}$ at the radius matching the median impact parameter (179 kpc). The implied value for $v_{cloud,s}$ would be about $10^2 ~km~s^{-1}$. 

Could this shocked gas produce the observed Si III, C IV, and O VI absorption-lines? First, we note that the radiative cooling time for the shocked gas will be $\sim 2 (Z_{\odot}/Z)$ Myr, where $Z$ is the cloud metallicity. For the typical metallicities of $\sim$ 0.3 solar inferred by Prochaska et al. (2017) and Werk et al. (2014), the cooling times are very short compared to the starburst lifetime. Shock models with velocities of $\sim 10^2$ km sec$^{-1}$  (e.g. Shull \& McKee 1979) can account for the observed ratios of the column densities of Si III, Si IV, and C IV. However, the ionic column densities of a single shock are small, and the observed values would imply that the line-of-sight through the CGM is intersecting tens of shocks. Shocks this slow do not produce significant O VI (e.g. Raymond 1979).  If O VI in the SB/CGM has the same physical origin as the lower ions, a range in shock speeds would be required (e.g. from $\sim 100$ to $\sim 200$~ km~s$^{-1}$). A range of shock velocities is to be expected, since a range of cloud densities in the pre-existing CGM is natural (Equation 6). 
 
An intriguing possibility is that the excess strength of the metal lines in the CGM of the COS-Burst galaxies could result from the destruction of CGM dust grains, releasing metals into the gas phase. This is a potentially important process: the amount of metals locked in dust grains in the CGM is similar to the amount present in the gas-phase in typical star-forming galaxies (M$\acute{e}$nard et al. 2010; Peeples et al. 2014). Calculations of grain destruction by shocks with speeds in the range we infer ($\sim$ 100 to 200 km s$^{-1}$) show that significant fractions of Si and C that could be liberated, with the fraction increasing with increasing shock velocity and decreasing grain size (Draine \& Salpeter 1979; B. Draine, private communication). While this possibility of dust destruction is not required in our model, it would have the advantage that the need for a bulk radial transport of gas-phase metals from the inner to outer CGM would not be required to explain all of the differences between the gas-phase metal distributions in the CGM of COS-Burst vs. normal star-forming galaxies.

\subsubsection{A Two-Phase CGM}

We now discuss how the above picture is modified if we add a diffuse volume-filling phase to the CGM. One immediate impact is that the timescale for the wind to affect the CGM can be significantly longer than in the case above. For simplicity, we consider a wind that propagates into a spherically-symmetric volume-filling CGM. This will create an expanding wind-blown bubble, a general problem that has been analyzed by Castor, McCray, \& Weaver (1975), Dyson (1989) and Koo \& McKee (1992). From inside-out, the structure of the wind-blown bubble will be:  the starburst (where the energetic wind fluid is created), a sonic point, a region of freely-streaming and adiabatically-cooling supersonic wind-fluid, an internal (reverse) wind shock, thermalized (shock-heated) wind fluid, a contact discontinuity, shocked CGM material, and an external (outer) shock being driven into the volume-filling CGM.

Observational constraints on the amount of hot volume-filling gas in the CGM of star-forming galaxies are rather poor. As summarized by Bland-Hawthorn \& Gerhard (2016) and Werk et al. (2014), there are estimates ranging from about $10^9$ to 10$^{11}$ M$_{\odot}$ for the halo of the Milky Way and other similar disk galaxies. The low densities (and hence low-inferred pressures) in the CGM absorption-line clouds found by Werk et al. (2014) for their sample would be more consistent with the low-end of this range. We note that the median value for the stellar mass in our sample is $\sim 2.2\times 10^{10}~M_{\odot}$, only about 40\% that of the Milky Way (Bland-Hawthorn \& Gerhard 2016). We will therefore simply parameterize the properties of the hot volume-filling phase for our sample by adopting a fiducial value of $M_{hot} = 10^{10}~M_{\odot}$.

We have shown above that the amount of kinetic energy supplied by an energy-driven wind is nearly two orders-of-magnitude larger than what would be required to move $10^{10}$ M$_{\odot}$ of gas outward in the halo potential well. We therefore ignore gravity. The densities in the region of the shocked wind and shocked CGM are so low that the radiative cooling times will be much longer than a Hubble time. We will therefore first consider an energy-driven bubble.

Both observations and theoretical models imply that the gas density in the CGM falls with radius roughly like $r^{-1}$ to $r^{-1.5}$ (Moller \& Bullock 2004; Miller \& Bregman 2013, 2015; Werk et al. 2014; Voit et al. 2016; Faerman, Sternberg, \& McKee 2017). Taking an initial radial density profile for the volume-filling phase of $n_{vf} \propto  1/r$, and following the self-similar solutions for energy-driven bubbles in Dyson (1989), the radius of the expanding bubble is given by:

\begin{equation}
r_{bubble}  = 86 \dot{E}_{43}^{1/4}  M_{hot,10}^{-1/4} t_8^{3/4} ~kpc
\end{equation}

Here $\dot{E }_{43}$ is the rate of kinetic energy injected by the starburst in units of $10^{43}$ erg sec$^{−1}$ (corresponding to a star-formation rate of 14 M$_{\odot}$ year$^{-1}$), M$_{hot,10}$ is the total mass of the volume-filling phase out to a radius of 200 kpc in units of $10^{10}$  M$_{\odot}$, and $t_8$ is the time since the starburst began given in units of $10^8$ years. 

The median starburst age is 280 Myr and the median kinetic energy injection rate implied by the median star-formation rate is $9 \times 10^{42}$  erg sec$^{-1}$  for our COS-Burst sample. This leads to an outer radius for the expanding wind-blown bubble in the halo of 182 kpc. Once reaching this radius, the outer shock speed ($dr_{bubble}/dt$) would be 480 km~s$^{-1}$. 

We can also consider a momentum-driven bubble, allowing for the possibility of radiative cooling being significant. In this case, the analysis in Dyson (1989) leads to:

\begin{equation}
r_{bubble} = 99 \dot{p}_{35}^{1/3} M_{hot,10}^{-1/3} t_8^{2/3}
\end{equation}

The median momentum-flux for the wind in the COS-Burst sample would be $6.6 \times 10^{34}$ dynes for $\alpha = 1$ and $\beta =$ 0.3 (see equation 4 above). This implies then that $r_{bubble} = 171$ kpc, and the current outer shock speed is $\sim 400$ \kms. Thus, the momentum-driven case yields very similar values for the bubble size and expansion speed. 

Given the highly idealized nature of our model and the uncertain mass of the volume-filling CGM, we regard these simple estimates as showing that it is plausible for a starburst-driven wind to affect the bulk of the CGM.

The interaction between this wind-driven bubble and pre-existing clouds in the CGM will depend on the location of the cloud. It can initially be overtaken by the outer shock driven into the CGM, it can then be compressed in the region of the thermalized wind fluid. In the interior region occupied by the free wind, the wind can accelerate and shock clouds, as described in the previous section.

\subsubsection{Wind-Stimulated Cloud Condensation}

In the model above, we have considered the interaction between a starburst-driven wind and pre-existing clouds in the CGM. An interesting alternative is the starburst-driven wind is instead actually creating the clouds seen in absorption by facilitating their condensation out of diffuse, thermally unstable gas in the CGM.

This idea has mostly been proposed and discussed in the context of the effects of AGN-driven outflows (jets) on the observed multi-phase gas in the cores of clusters of galaxies (Li \& Bryan 2014; Voit et al. 2016). In this model, the AGN-driven outflow uplifts diffuse ambient gas. This uplifted gas cools adiabatically, thereby shortening its radiative cooling time and promoting the development of thermal instabilities and the formation of cold condensates (clouds). As noted by Voit et al., this idea could be generalized to the case of the CGM and outflows driven by feedback from massive stars.

In the context of this paper, this idea is attractive because it circumvents the difficulty in preventing wind-accelerated pre-existing clouds in the CGM from being destroyed by hydrodynamical instabilities before they can be accelerated to high velocities or transported over significant distances (see Heckman \& Thompson 2017 and references therein).

\section{Conclusions}

The circum-galactic medium (CGM) represents the potential source of gas to fuel the future growth of the galaxy through accretion and subsequent star-formation. In this paper we have investigated the effects of the energy and momentum released by a starburst on the CGM, in order to understand the role of such feedback in the evolution of galaxies. We have used Hubble Space Telescope’s (HST) Cosmic Origins Spectrograph (COS) to measure the far-UV spectra of background quasars along lines-of sight passing within $\sim$ 230 kpc of 17 low-redshift starburst and post-starburst galaxies (the COS-Burst sample). We have detected the CGM in absorption with Ly$\alpha$ in 100\% of the possible cases, with Si III 1206.5 in 47\% , and with C IV 1548.2 in 44\%. In six cases we accessed the O VI 1031.9 line detecting it in three cases (50\%). We have only upper limits for Si II 1260.4, C II 1334.5, and Si IV 1393.8.

We have used archival HST COS data to define control samples of normal star-forming galaxies selected to have the same range in stellar mass ($\sim 10^{10}$ to 10$^{11}$ M$_{\odot}$) and impact parameter as the COS-Burst sample. We have then compared the properties of the CGM in the COS-Burst and control samples. We found:

\begin{itemize}
\item[1.] The Ly$\alpha$, Si III, and C IV absorption-lines are significantly stronger in the CGM of the COS-Burst galaxies. This comparison was done as a function of normalized impact parameter ($\rho/R_{vir}$). We note that this difference pertains to the outer CGM ($\rho/R_{vir} > 0.5$), since have few COS-Burst sightlines in the inner CGM.
\item[2.] Both the (non-parametric) full-width at half maximum ($FWHM$) and the velocity displacement of the individual Ly$\alpha$ lines with respect to the galaxy systemic velocity ($\Delta v$) are significantly larger for the COS-Burst galaxies. The stacked Ly$\alpha$ absorption-line profile for the COS-Burst sample is roughly two times wider than the value expected if the clouds are moving through the CGM at the halo virial velocity (FWHM = 424 vs. 214 \kms).
\item[3.] The ratios of the equivalent widths of the Si III and C IV lines to those of Ly$\alpha$ are larger in the CGM of the COS-Burst galaxies (by an average of 0.4 dex).
\item[4.] We conclude that the amount of metals is enhanced and the dynamical state of the (outer) CGM is significantly different in the COS-Burst sample.
\end{itemize}

The detected metal absorption-lines are not usually saturated in the COS-Burst sample ($<\tau> \sim 1$). Using the measured impact parameters and column densities we infer masses of M$_{Si II, III, IV} = 4.0$ to 10.0 $\times 10^5$ M$_{\odot}$, M$_{C IV} = 2.4 \times 10^6$ M$_{\odot}$, and M$_{OVI} = 5 \times 10^6$ M$_{\odot}$ for the outer CGM ($\rho =$ 50 to 200 kpc). 

We next considered the causal relationship between the unusual properties of the CGM and the presence of a starburst. The COS-Burst galaxies are mostly normal late-type galaxies (not byproducts of recent major mergers). More generally, we argued that an inwardly directed connection (e.g. an unusual CGM leads to the triggering of a starburst) was unlikely. Given the typical starburst ages (median of 280 Myr) and impact parameters (median of 179 kpc) the required minimum velocity of the inflow would be $\sim$ 630 km s$^{-1}$, compared to a median halo circular velocity of only 129 km s$^{-1}$. We also explored the possibility that radiation pressure from the starburst could accelerate CGM clouds outward. Given the very low estimated dust optical depths of the clouds, this fails by orders-of-magnitude.

We therefore considered the possible impact of a galactic wind driven by the collective momentum and/or energy supplied by supernovae and stellar winds in the starburst. We showed that (in the absence of a volume-filling CGM phase) a starburst-driven galactic wind could accelerate and shock-heat the types of clouds that are observed in the CGM of normal galaxies, and do so over the full range of observed impact parameters. We noted that it was possible these shocks could also liberate metals from CGM dust grains. If there is a significant volume-filling phase (as well as clouds) we showed that energy and/or momentum supplied by the starburst-driven wind could inflate a CGM-scale wind-blown bubble with the required size, on the required time-scale. We also suggested that a starburst-driven wind could facilitate the creation of new clouds as adiabatically cooled uplifted diffuse gas in the CGM became thermally unstable.

While starbursts as strong as our sample are atypical in the present-day universe, they have specific star-formation rates ($\sim 10^{-9}$ yr$^{-1}$) that are typical of normal star-forming galaxies at redshifts of $\sim$ 0.5 to 3, the epoch during which $\sim$80\% of the present-day cosmic stellar inventory was created (Madau \& Dickinson 2014). Indeed, absorption-line probes of the CGM around normal star-forming galaxies at $z \sim 2$ to 3 also show radial distributions of Ly$\alpha$ and C IV absorption-lines extending out to impact parameters of $\sim$ 300 and $\sim$ 100 kpc respectively (Steidel at al. 2010). The physical picture we have proposed for the CGM of present-day starburst/post-starburst galaxies should be broadly relevant to the evolution of galaxies over cosmic time. Our data also yield new information about both starburst-driven winds and the CGM itself, as they show how the CGM responds to the injection of energy and momentum supplied by a wind on global scales.
We believe these new data can provide a valuable benchmark for future numerical simulations of the effects of feedback on galaxy evolution (e.g. Simpson et al. 2015; Muratov et al 2015; Li et al. 2017; Hopkins et al. 2017).

\vspace{.5cm}
\acknowledgements 

This work is based on observations with the NASA/ESA Hubble Space Telescope, which is operated by the Association of Universities for Research in Astronomy, Inc., under NASA contract NAS5-26555. TH and SB were supported by grant HST GO 13862. VW acknowledges the support of the European Research Council via the award of a starting gant (SEDMorph: P.I. V. Wild).

This project also made use of archival SDSS data. Funding for the SDSS and SDSS-II has been provided by the Alfred P. Sloan Foundation, the Participating Institutions, the National Science Foundation, the U.S. Department of Energy, the National Aeronautics and Space Administration, the Japanese Monbukagakusho, the Max Planck Society, and the Higher Education Funding Council for England.  The SDSS Web Site is http://www.sdss.org/. The SDSS is managed by the Astrophysical Research Consortium for the Participating Institutions. The Participating Institutions are the American Museum of Natural History, Astrophysical Institute Potsdam, University of Basel, University of Cambridge, Case Western Reserve University, University of Chicago, Drexel University, Fermilab, the Institute for Advanced Study, the Japan Participation Group, Johns Hopkins University, the Joint Institute for Nuclear Astrophysics, the Kavli Institute for Particle Astrophysics and Cosmology, the Korean Scientist Group, the Chinese Academy of Sciences (LAMOST), Los Alamos National Laboratory, the Max-Planck-Institute for Astronomy (MPIA), the Max-Planck-Institute for Astrophysics (MPA), New Mexico State University, Ohio State University, University of Pittsburgh, University of Portsmouth, Princeton University, the United States Naval Observatory, and the University of Washington.

TH thanks the Kavli Institute for Theoretical Physics, the Simons Foundation, and the Aspen Center for Physics for supporting informal workshops on galactic winds, feedback, and the circum-galactic medium during 2014, 2015, and 2016. These provided important input for this paper.  Conversations at these workshops with Romeel Dav{\'e}, Crystal Martin, Norm Murray, Ralph Pudritz, Eliot Quataert, Brant Robertson, Chuck Steidel, Todd Thompson, and Ellen Zweibel were especially helpful. He also thanks Greg Bryan, Mark Voit, and Brian O`Shea for interesting discussions of the possible importance of wind-driven precipitation.

{\it Facilities:}  \facility{Sloan ()} \facility{COS ()}

\end{document}